\documentclass[12pt]{article} 
\RequirePackage[colorlinks,citecolor=blue,linkcolor=blue,urlcolor=blue,pagebackref]{hyperref}

\usepackage{amsmath,amsfonts,esint}
\usepackage{amsmath,amstext,rotating}
\usepackage{amsfonts,amssymb,graphics,xspace,endnotes}
\usepackage{lineno}
\usepackage{epsfig}
\usepackage{srcltx}
\usepackage{amsthm}
\usepackage{geometry}
\usepackage{subfig}
\usepackage{setspace}
\usepackage{graphicx}
\usepackage{xcolor}	
\usepackage{verbatim}
\usepackage{natbib}

\newcommand{\mcF}{{\mathcal F}}

\newcommand{\mcZ}{{\mathcal Z}}

\newcommand{\real}{{\mathbb R}}

\def \PP {\mathbb{P}}
\def \EE {\mathbb{E}}

\def \cB {\mathcal{B}}

\def \cV {\mathcal{V}}
\def \rT {\mathbb{T}}
\def \rbT {\mathbb{\bf T}}
\def \rbX {\mathbb{\bf X}}

\theoremstyle{plain}
\newtheorem{prop}{Proposition}

\newtheorem{assumption}{ASSUMPTION}
\newtheorem{theorem}{THEOREM}
\newtheorem{lemma}{LEMMA}
\newtheorem{corollary}{COROLLARY}

\theoremstyle{remark}
\newtheorem{definition}{DEFINITION}

\DeclareMathOperator{\Argmin}{Argmin}

\usepackage{natbib}
\setcitestyle{sort&compress}

\AtBeginDocument{}

\hypersetup{
    colorlinks,
    linkcolor={blue!75!black},
    urlcolor={blue!75!black},
    citecolor={blue!75!black},
}

\begin{document} 

\title{Privacy-aware identification\thanks{We thank C. Dwork for conversations 
and discussions with us which ultimately directed us to 
study differential privacy.
The current draft of the paper 
has greatly benefited from the insightful
comments from  J. Hotz, C. Manski, R. Moffitt, F. Molinari, 
 J. Pepper, A. Smith and C, Tucker as 
well as seminar participants at various universities. Support from STICERD and the NSF is gratefully acknowledged.}} 
\author{Tatiana Komarova\thanks{Faculty of Economics, University of Cambridge.  tk670@cam.ac.uk.}  \and
Denis Nekipelov\thanks{Department of Economics, University of Virginia.  denis@virginia.edu.}} 

\date{\today \\ \vspace{.5cm}}

\maketitle



\begin{abstract}
The paper redefines econometric identification under formal privacy constraints, particularly differential privacy (DP). Traditionally, econometrics focuses on point or partial identification, aiming to recover parameters precisely or within a deterministic set. However, DP introduces a fundamental challenge: information asymmetry between researchers and data curators results in DP outputs belonging to a potentially large collection of differentially private statistics, which is naturally described as a random set. Due to the finite-sample nature of the DP notion and mechanisms, identification must be reinterpreted as the ability to recover parameters in the limit of this random set.  In the DP setting this limit may remain random which necessitates new theoretical tools, such as random set theory, to characterize parameter properties and practical methods, like proposed decision mappings by data curators, to restore point identification. 
We argue that privacy constraints push econometrics toward a broader framework where randomness and uncertainty are intrinsic features of identification, moving beyond classical approaches. By integrating DP, identification, and random sets, we offer a privacy-aware identification.

\vspace{0.05in}

\begin{description}
\item[Keywords:] Differential privacy, random sets, identification, average treatment effect
\end{description}
\end{abstract}
\newpage

\section{Introduction}\label{intro}

The era of broadly accessible digital data has transformed economics, enabling precise estimation of causal effects, policy impacts, and behavioral patterns. Yet, this power comes with a cost as sensitive datasets ( such as those revealing incomes, health records, or employment histories) pose significant privacy risks. When a census links individual earnings to addresses or a survey exposes personal choices, the potential for harm grows. Differential privacy (DP), a rigorous framework from Computer Science, addresses this by adding controlled noise to data outputs, ensuring that no individual’s information can be reverse-engineered.  But, as we show in this paper, this protection challenges the traditional approach to a cornerstone of econometrics: identification. 

This paper argues that the advent of formal privacy guarantees, epitomized by differential privacy, introduces a new reality in how we should approach identification in sensitive data and privacy-protected outputs. In  a nutshell, when a researcher does not have access to raw data but instead is only given a privacy-protected output, that creates informational asymmetries between data set handlers (subsequently called data curators) and researchers. This is because a released output depends on the noise infused to deliver privacy guarantees with this noise not known to a researcher. Moreover, a researcher typically does not have a full knowledge of the parameters of that noise distribution (ultimately linked to desirable privacy level) or the algorithms by  which the noise has infused, or is unable to confirm that the statistic on which the randomization of the output is based is even a suitable approach for given data (this is particularly of concern when the statistic itself depends on the tuning parameters). Given these informational asymmetries, an \textit{honest} researcher would have to treat a given randomized statistic as an element of a potentially large collection of statistics naturally described as a random  set. 

Our paper thus proposes a paradigm shift in 
econometrics, leveraging random set theory  to redefine identification in the presence of differential privacy. Since the parameters of interest are now elements of random sets, their properties are governed by a containment functional akin to a probability measure. We show  that even with infinite data, outputs may not converge to a single value or a deterministic set. The traditional notions of point and partial identification consequently often fail. While the need for a probabilistic approach that characterizes parameters as elements of random sets  introduces complexity to the analysis, it  also unlocks opportunities. We propose that data curators, who control access to sensitive datasets, can restore point identification by applying decision mappings which are principled rules that select a single estimate from the random set based on its geometry. This transforms the curator’s role from a gatekeeper to a collaborator in scientific discovery, ensuring that privacy-protected data remains useful.

In a nutshell, our work builds on three strands of research: differential privacy, econometric identification, and random set theory, combining them into a new framework for privacy-aware identification.




Section \ref{sec:DPreview}  reviews the notion of DP, introduced by \citet{dwork_etal:06} with the purpose of providing a rigorous framework for protecting sensitive data.
 It has been a breakthrough   concept
and there is a broad consensus in the Computer Science community that DP is the only existing privacy concept that can help safeguard the output of computing on data from a broad range of disclosure risks, prohibiting re-identification of individuals in the data or their characteristics. Informally, the core idea of DP is based on the randomization of the output of statistics computed from the original datasets. The infused noise must guarantee that the probability distribution of the randomized statistic can only vary within a given small range if any observation in the data is removed or altered. This guarantee must hold for any realization of the dataset, even those that may occur with very low 
probability.

Looking at it from another angle, DP subsumes a structured model for the evaluation of data statistics with three main components: the dataset, the data curator, and the researcher. The researcher is interested in the estimation of a specific parameter from the data. The data curator fully controls the data and does not allow the researcher to interact with the data directly. Instead, the data curator herself forms an  output that is computed from the dataset and infused with noise to provide a concrete DP guarantee.

Section \ref{sec:DPreview} discusses important aspects of the DP notion, such as e.g. the privacy budget allocation to the entire dataset which particularly makes it impossible  for a researcher to obtain several different randomized outputs for a requested statistic. Another important issue we emphasize in that section is a deep informational asymmetry between the data curator and the researcher. While some level of informational asymmetry is inherent to any privacy-preserving system (after all, the very purpose of such systems is to withhold certain information from the researcher) our findings reveal a more problematic form of asymmetry. Specifically, the lack of identification arises due to two key factors:
(i) researchers are often provided only vague (sometimes none) information about the parameters of the DP mechanism; and
(ii) researchers cannot verify assumptions about the underlying data distribution, which are critical for ensuring the validity of standard statistical procedures. Factor (i) is relatively intuitive. Factor (ii) is particularly problematic in practice. Many statistical methods rely on distributional assumptions, such as the existence of higher-order moments (e.g., variance, skewness) or sub-Gaussian tails. In non-private settings, researchers typically validate these assumptions by inspecting plots or histograms, checking for familiar shapes such as the bell curve indicative of normality. DP, however, severely limits such exploratory analysis. As a result, researchers risk applying statistical tools that do not align with the actual data distribution. This mismatch can lead to analyses based on incorrect assumptions, This is particularly of concern in analyses involving nuisance parameters.

Even though Section \ref{sec:DPreview} does not contain any novel results or notions, it perhaps gives a fresh perspective at some of the issues related to the use of DP which have been overlooked to some extent by the Computer Science literature and that will ultimately be essential for our identification approach. 

Section \ref{sec:privacyaware} presents the various components of our privacy-aware identification approach. It begins with a high-level overview of the framework, followed by a discussion of how informational asymmetries between the data curator and the researcher inevitably lead the researcher under an \textit{honest} approach to interpret any given randomized output as part of a broader collection of possible randomized statistics, conditional on all information available to them. This perspective is naturally captured by treating each randomized output as a selection from a \textit{random set}.

Section \ref{sec:identification_formal} describes the fundamental properties of the statistics that are of interest to the researcher. We formulate  that these statistics possess desirable regularity properties and, in the absence of privacy-protection mechanisms, would provide a sound framework for statistical inference on the parameter of interest.

Section \ref{ex:samplemean2} provides an example of a weighted mean as the parameters of interest. This example illustrates the dependence of the estimator/statistic of interest on nuisance parameters and the informational asymmetries driven by that. We also take this example to illustrate the most common practical DP mechanisms for delivering DP outputs and illustrate their working  and properties in the context of this example. 
This example is used throughout the paper to illustrate various components involved into our privacy-aware identification analysis.


Section \ref{sec:directID} provides a detailed treatment of the \textit{direct privacy-aware identification} approach. This framework builds upon the researcher’s available information regarding the strength of privacy guarantees, the specific DP mechanisms employed by the data curator (this includes consideration of the tuning parameter behavior), but it assumes no further involvement from the curator. In other words, the curator performs her traditional role without explicitly accounting for the impact of mechanism-induced noise on parameter identification.

Because practical DP algorithms are finite-sample based, our notion of identification must rely on a limiting concept for random sets. Owing to the independence properties of mechanism noise, we argue that the appropriate limit is the weak limit. We show that informational asymmetries can (and often do)  cause estimators that would otherwise be consistent in the absence of privacy constraints to converge weakly to a non-deterministic random set. Consequently, under DP, parameters of interest may be neither point- nor partially identified, even with an infinite amount of data (a direct point identification of the parameter of interest from DP outputs is only possible if the limiting random set has a degenerate distribution at the true parameter value).
Consequently, identification analysis must rely on tools from the theory of random sets developed in \cite{molchanov2005}, \cite{BeresteanuMolinari08}, \cite{beresteanu2011a}, \cite{beresteanu2012},   \cite{MolchanovMolinari18}.\footnote{Classical tools of partial identification, such as those formulated and in \cite{manski:2003}, \cite{manski:2002}, among many others, are not applicable.}  In particular, central to us is the concept of the Choquet capacity, which generalizes the notion of a probability measure. This capacity fully characterizes the distribution of the random set and thus encapsulates all information about the population distribution that can be inferred from the DP output. We discuss why the concept of selection (Aumann) expectation or Vorob'ev's expectation which are traditional in the identification literature that utilizes random sets are not suitable within the DP setting due to privacy budget considerations. 

Thus, Section \ref{sec:directID} addresses the issue of  the fundamental tension that exists between parameter identification and the level of noise  required by differential privacy (DP) guarantees as well as the degree of informational asymmetry. Importantly, our results are general: they do not depend on any specific implementation of a DP mechanism. As long as the DP output satisfies our mild regularity conditions, the identification challenges we highlight will still apply.

Section \ref{sec:smoothness} defines a class of DP estimators/statistics that are approximately separable  
into a data-dependent component  and a mechanism noise component, 
with a negligible residual term. We refer to them as \textit{smooth} estimators. 
This structure arises naturally in many DP mechanisms (to the best of our knowledge, all most common practical DP mechanisms satisfy this property), which is illustrated in this section too. Smoothness enables a tractable analysis of the weak limit of the random set of DP estimators,  
which is central to assessing identifiability under DP,

Section \ref{sec:datacurator} develops an approach for restoring point identification in DP estimation when the limiting set of DP outputs is not a single true parameter value, which is a situation that frequently arises in applied settings. For such cases, the section introduces an \textit{active collaboration framework} within our \textit{privacy-aware identification} approach, in which the data curator plays a constructive role by using knowledge of the distribution and geometry of the DP outputs to select a unique, reproducible estimate. The key idea is a decision mapping that chooses one point from the random set of DP estimators using a strongly convex criterion. This procedure ensures transparency, preserves formal privacy guarantees, and yields consistent estimation of the target parameter. The section further explores practical ways to construct suitable decision mappings, particularly when the true parameter can be represented as a weighted integral (or geometric centroid) of the limiting random set, allowing implementations based on surface-integral representations. Together, these results demonstrate how structured, informed actions by the data curator can recover point identification within the DP framework without sacrificing privacy.

Section \ref{sec:bayes} shifts the focus from identification to inference, addressing situations where the data curator does not collaborate to ensure point identification and the researcher observes only a single DP output. In this case, the true parameter cannot be identified, but meaningful inference is still possible. The section develops a Bayesian framework that uses the distribution of the DP output, together with prior information, to construct credible regions quantifying uncertainty about the parameter of interest. This approach provides a practical and coherent method for inference when frequentist confidence intervals are infeasible under privacy constraints, extending ideas from Bayesian analysis of weak or partial identification to the DP setting.

Section \ref{sec:appl} illustrates the practical implications of our \textit{direct identification approach} in Section \ref{sec:directID} by applying it to the estimation of average treatment effects (ATE).

Our first applied illustration uses 
 inverse propensity score (IPS) estimators. The analysis shows how DP affects identification in this familiar econometric setting. Namely, the need to bound or trim propensity scores to ensure privacy can distort identification and lead to non-degenerate limiting distributions in the direct identification approach. The section characterizes the resulting regimes under which privacy noise dominates or vanishes, demonstrating that DP can fundamentally alter identification properties even in standard treatment-effect models.

Our second applied illustration examines how DP affects identification in regression discontinuity designs (RDD). It shows that the loss of point or partial identification under privacy constraints stems from the high sensitivity of nonparametric RDD estimators to individual observations, compounded by informational asymmetries arising from the researcher’s limited knowledge of the exact DP constants or mechanisms. Because privacy noise must accommodate this sensitivity, it does not vanish asymptotically, producing non-degenerate limiting random sets in the direct identification approach. 

In these applied illustrations in the presence of informational asymmetry, identification can be restored only through active collaboration with the data curator.

\paragraph*{Literature review and the place of our paper in the literature}

As already mentioned above, DP originated in Computer Science, where a large body of work has developed formal mechanisms for releasing randomized statistics that satisfy provable privacy guarantees (see, e.g., \cite{dwork_etal:06}, \cite{mcsherry:07}, \cite{nissim:07}, \cite{dwork2014algorithmic}). These methods have since been integrated into a wide range of machine learning procedures, producing privacy-preserving versions of classification, regression, and empirical risk minimization algorithms (e.g., \cite{chaudhuri2009privacy}, \cite{rubinstein2009learning}, \cite{friedman2010data}, \cite{chaudhuri2011differentially}, \cite{kifer2012private}, \cite{bassily2014differentially}, \cite{AbadiChu_etal16}, among many others).

In applied work, there are two conceptually distinct ways in which DP mechanisms may be implemented. In the first, which we focus on in this paper, the data curator produces a noisy version of the exact statistic that the researcher wishes to compute, thereby directly privatizing the target estimand. In the second, the curator releases a collection of predefined statistics (e.g., aggregated means or quantiles of key variables) each made differentially private. In that case, the researcher must construct the desired statistic from the released quantities. The DP guarantee remains valid in this second setting by virtue of the post-processing theorem (see \cite{dwork2014algorithmic}), which ensures that any function of DP outputs preserves the privacy guarantee.

From an econometric perspective, most of the DP literature is limited by its focus on the privacy properties of released outputs rather than on the identification and inference of underlying population parameters. While the DP and broader statistical disclosure control literatures have recognized that privacy noise may degrade statistical accuracy, their main focus has been on the so-called \textit{privacy–utility trade-off} which looks at how increasing privacy protection raises the mean-squared error or reduces predictive performance (see, e.g., \cite{lindell2000privacy}, \cite{karr:2006}, \cite{brickell2008cost}, \cite{woo:2009}).

This perspective overlooks a central concern in econometric analysis which is related to the privacy noise not merely increasing variance but potentially fundamentally altering the identification structure of the model. A long line of research in econometrics (e.g., \cite{horowitz:1995}, \cite{molinari2008partial}) shows that stochastic contamination of data can lead to loss of point identification that persists even in infinite samples, necessitating partial identification methods. From this standpoint, DP mechanisms can change the limiting behavior of estimators in a way that is qualitatively different from the standard ``variance inflation'' intuition prevalent in the Computer Science literature.

Our results also relate to existing impossibility results for DP (e.g., \cite{DSSU17}, \cite{JOU20}), though the source of the difficulty is different. Those papers typically derive statistical lower bounds for finite-sample estimation under privacy constraints. In contrast, our analysis focuses on the asymptotic identification of structural parameters when estimators are privatized, and develops a framework for restoring point identification through structured interaction between the researcher and the data curator.

It is also important to clarify what this paper is \textbf{not} about. We do not design bespoke DP mechanisms tailored to individual parameters of interest. Such an approach may be suitable for large, one-time data releases (such as the Decennial Census) but is less realistic for ongoing empirical work. Instead, we consider a setting in which the data curator applies a generic DP mechanism to all data requests, regardless of the analysis being conducted. Nor do we rely on replacing model-driven estimators with robust alternatives that happen to satisfy DP assumptions, since the econometric estimand is typically dictated by the structural model and cannot be freely altered.

In this sense, our paper bridges the DP and econometrics literatures by showing that privacy guarantees can reshape identification theory and by developing a framework in which active collaboration between the researcher and data curator can recover consistent, reproducible inference while maintaining formal privacy protection.

\section{Differential privacy concept}
\label{sec:DPreview}

In this section, we first review the notion of DP and then discuss its aspects that are of particular importance for our privacy-aware identification. This section does not present any new results but perhaps provides some fresh perspective on the interpretation of DP key components.


 The Computer Science literature generally envisions a researcher (with no direct access to the data) sending a \textit{query} to the data curator (who has access to the data)  requesting a statistic of interest which we will denote as $\theta(\PP_N,0)$. This statistic may represent a good estimator of the parameter of interest. The data curator then decides on the choice of degree of privacy protection (formalized in privacy constraints in Definition \ref{def:privacy} below) and a particular mechanism to instead deliver a statistic $\theta(\PP_N, \nu_N)$ with the chosen degree of privacy protection. 

\begin{definition}[\textit{$(\epsilon, \delta)$-differential privacy,  \cite{dwork:2006}}] 
\label{def:privacy}
A randomized statistic $\theta(\PP_N,\nu_N)$ is $(\epsilon, \delta)$-differentially private if for any two empirical measures $\PP_N$ and $\PP'_N$ over $N$ support points and 
differing in exactly two support points, we have that for all measurable sets $A$ of possible outputs the following holds: 
\begin{equation} \label{epsdelta_def} \PP_{\nu_N}\left[\theta(\PP_N,\nu_N) \in A \right] \leq e^{\epsilon} \PP_{\nu_N}\left[\theta(\PP_N',\nu_N) \in A \right] + \delta,   
\end{equation}
where $\epsilon > 0$, $\delta \in [0, 1)$ are \textit{privacy parameters} (sometimes we will refer to them as \textbf{privacy constants}), and probabilities are taken over randomness in $\nu_N$. 

In addition, if $\delta=0$, then the estimator $\theta(\PP_N,\nu_N)$ is referred to as $\epsilon$-differentially private.
\end{definition}	

$(\epsilon, \delta)$-differential privacy is also known also known as ``approximate differential privacy". Notation $\PP_{\nu}(\cdot)$ is used to emphasize that differentially private randomized statistic $\theta(\PP_N,\nu_N)$ is based on the distribution of random element $\nu_N$ while the distributions of two adjacent datasets $\PP_N$ and $\PP_N'$ are fixed.  The randomized estimator $\theta(\PP_N,\nu_N)$ 
in Definition \ref{def:privacy} ensures that information regarding individual data entry cannot be reverse engineered from its values.  DP requires that any statistic must be independently randomized such that its distribution over the introduced randomness is “not very sensitive” to changes in individual observations in the sample.

\paragraph*{Important aspects of Definition \ref{def:privacy}} We highlight two important implications of Definition \ref{def:privacy}. First, while Definition \ref{def:privacy} allows the distribution of $\nu_N$ to be tailored to general characteristics of the \textit{population} data distribution reflected in $\PP_N$ (such as the number of variables or their support) and allows dependence on the sample size $N$, it explicitly prohibits dependence on the specific observed values that constitute $\PP_N$ (e.g., the realized sample support), see \cite{dwork:2008} Remark 1.2 (page 3). Second, the privacy guarantee must be maintained for every possible realization of the datasets $\PP_N$ and $\PP_N'$ that differ by exactly one support point, regardless of how small the probability of such realizations may be. Together, these two features underscore the strong privacy protections inherent to DP. However, as evident from our discussion later, these same properties that make it a strong privacy notion can also contribute to identification challenges for parameters in differentially private versions of certain important econometric models. 

For readers not very familiar with the DP concept, it is worth mentioning a specific case $\delta=0$. Then inequality (\ref{epsdelta_def}) can be rewritten in terms of a familiar log-likelihood ratio: 
\begin{equation*}\log \frac{\PP_{\nu}\left(\theta(\PP_N,\nu) \in A \right)}{\PP_{\nu}\left(\theta(\PP_N',\nu) \in A \right)} \leq {\epsilon},   
\end{equation*} with the constant $\epsilon$ capturing the maximum privacy loss. The differential privacy requirement with $\delta=0$ is a rather strong one as it puts a bound on the privacy loss even for very unlikely events (when $\PP_{\nu_N}\left(\theta(\PP_N,\nu_N) \in A \right)$ or $\PP_{\nu_N}\left(\theta(\PP_N',\nu_N) \in A \right)$ is extremely small).  With $\delta>0$,  ``bad'' events from the perspective of the log-likelihood ratio are allowed to happen but the probability of such is bounded by $\delta$. For further discussion, see \cite{dwork2014algorithmic}.

\paragraph*{Important aspects not covered by Definition \ref{def:privacy}}
Definition \ref{def:privacy} gives a criterion for a statistic to be considered differentially private but is silent about some other important features of a privacy-protecting process.  

First, it does not constrain a data curator to use any particular mechanism to achieve DP.  The Computer Science literature has suggested some practical mechanisms for delivering DP statistics for general queries and we review some of them in our examples later in the paper. Overall, a DP mechanism consists of both the form of the \textit{functional} $\theta(\PP_N,\nu_N)$ which will refer to as an \textit{algorithm} and the injected \textit{noise distribution}  $\nu_N$. 

Second,  Definition \ref{def:privacy} is subtle  about  how much knowledge about the mechanism and privacy constraints is communicated by the  data curator to the researcher. A researcher is viewed as a potential adversary who may threaten privacy so  $\nu_N$ together with the knowledge of the algorithm $\theta(\cdot, \nu_N)$ is not revealed as then  the researcher may be able to uncover $\theta(\PP_N,0)$ of interest violating the criterion in Definition \ref{def:privacy}. Some knowledge about the mechanism that complies with Definition \ref{def:privacy} may, however, be communicated to the researcher. The extent of communicated knowledge differs across data curators. Some may release  no knowledge, some may release the form of the functional $\theta(\cdot,\cdot)$, some may release privacy parameters $(\varepsilon,\delta)$, and a combination of these. Thus, from the perspective of a researcher there are multiple algorithms which can produce a DP randomized statistic, for instance, by adding noise to the non-randomized statistics, resampling it or use other methods such as quasi-Bayesian sampling which is not directly tied to a particular non-private statistic. Moreover, each method requires calibration of parameters such as the variance of the added noise or number of subsamples leading to a continuum of statistics conforming to the DP criterion. Note that  even knowledge of  $(\varepsilon,\delta)$ only yields the
upper bound guaranty for DP of a given estimated output due to the requirement $\leq$ rather than the requirement $=$ in Definition \ref{def:privacy}. This implies that ``effective'' $(\tilde{\varepsilon}, \tilde{\delta})$ under which we have $=$ in the criterion in Definition \ref{def:privacy} for some sets $A$ may actually be coordinate-wise smaller than $(\varepsilon,\delta)$ communicated by the data curator. Another cause of informational asymmetry stems from multiple ways to set tuning parameters
which are required when the  desired statistic $\theta(\PP_N),0$ itself  depends on the tuning parameters (as nonparametric or high-dimensional estimations  do) which then naturally in incorporated into a DP algorithm. 
The researcher is, of course, unable to engage in an exploratory data analysis to ensure that the necessary properties of the parameter of interest  $\theta(\PP_N,0)$ even in the absence of the mechanism noise $\nu_N$. Thus, with DP outputs a researcher has to be honest about the lack of knowledge regarding how tuning parameters interact with privacy constraints in a randomized statistic which in itself affects the statistical properties of a randomized statistic. We illustrate all the aspects mentioned here in our Example in Section \ref{ex:samplemean2}.



\paragraph*{Multiple queries, privacy budget }

One {might mistakenly think that} an effective way 
to mitigate the randomness introduced by $\nu_N$ is to consider repeatedly sampled
versions of the randomized statistic
$\theta(\PP_N,\nu_N).$ This {thinking is erroneous} because the described scheme is impossible due to the 
DP requirement of considering {\it the 
entire set} of those statistics 
{\it as a single vector-valued statistic.} This 
DP requirement is often implemented by
allocating a {\it privacy budget}
$\epsilon^*$ to {\it all} statistics
that would ever be 
evaluated from a given dataset.
Then each statistic would be allocated
a fraction of that $\epsilon^*$ for the corresponding $(\epsilon,\delta)$ - differentially private output (desirably, with $\epsilon \ll \epsilon^*$) and the more statistics are evaluated
from the data, the less of the
the privacy budget $\epsilon^*$
will be eventually allocated to the remaining queries and, consequently, the more noise
will have to be added to those
queries. To the best of our knowledge, the current DP practice
to avoid the unnecessary erosion
of the privacy budget is to allow
each unique statistic of the 
data to be evaluated exactly once. With this rule, each subsequent evaluation 
of that function would use 
the same noise that 
would be added to ensure
differential privacy and output
the value identical to that in
its first ever evaluation.The consequences of
this feature are discussed later.

\section{Privacy-aware identification} 
\label{sec:privacyaware}

Having reviewed the DP concept, we now turn to the econometric aspect of this paper moving towards our main research question of approaching identification within the DP paradigm.

Before proceeding with details of the formal framework, we present a high-level overview of our approach. {In line with the general econometric and statistical research}, the object of interest in this paper is parameter $\theta_0$ which can be expressed as a functional of the population distribution of the vector of random variables $Z.$ In finite samples this parameter can be estimated using various statistics of the data. Our focus is the identification of parameter $\theta_0$ when any such statistic produced from the data must be differentially private. 

The  considerations in Section \ref{sec:DPreview} that relate to informational asymmetries (either driven by possibility of multiple algorithms or unknown sharp privacy constraints, or  researcher's lack of knowledge how tuning parameters interact with privacy constants)  justifies  the need for a researcher in an \textit{honest}  approach  to the analysis of DP outputs to treat  each DP output as part of a potentially large collection of differentially private statistics. More formally, in an honest approach to identification each DP output needs to be treated as  a realization of a \textit{selection} from the random set which is compact subset of a metric space and comprises all DP statistics possible under the information available to the researcher.


The identification analysis must then, first, consider this whole random set, and, second,  be 
performed in the limit because the DP 
criterion only applies to finite samples and cannot 
be applied directly to some functional of the 
population distribution.\footnote{Similar issues applies to statistical disclosure limitation approaches like $k$-anonymity addressed in \cite{KNY:18}} Thus, the identification of 
target parameter $\theta_0$ boils down to the 
analysis of the behavior of the limit of the collection of differentially private statistics 
as the sample size increases and the empirical distribution of the data converges to the population distribution of the underlying random variable $Z$.  In other words, identification of $\theta_0$ must be defined as a property of the limit of a sequence of random sets as the sample size approaches infinity. 


We now give a step-by-step coverage of our privacy-aware identification approach. 

\subsection{Statistics/estimators of interest}
\label{sec:identification_formal}


We consider a sequence of statistical experiments 
indexed by the sample size $N$ ($N \rightarrow \infty$ along
this sequence), where for each $N$ we generate an i.i.d. sample $\{z_i\}^N_{i=1}$ from the joint distribution
of $d$-dimensional random vector $Z$ leading to empirical distribution $\PP_N$. We assume that the parameter of interest $\theta_0$ is in the interior of 
$p$-dimensional convex compact parameter space $\Theta\ \subset \real^p$.

In line with the DP setting, we then consider randomized statistics $\theta(\PP_N,\nu_N) \in \Theta$ which are functionals
of the empirical distribution of the data and independent random element
$\nu_N$ is the key feature of the 
randomized statistics that allows
it to provide a differential privacy guarantee.




Assumption \ref{operators} below gives a formal description of the class of  statistics/estimators that will be building blocks of a random set considered  by the researcher in an honest approach.

\begin{assumption}\label{operators}
The class of randomized statistics $\mathcal T$ producing $(\varepsilon_N, \delta_N)$-differentially private estimators 
for the target parameter $\theta_0$
is associated with operators $\theta(\cdot,\cdot)$ such that:
\begin{itemize}
 \item[(i)] For
 each $\nu \in \cV,$  $\theta(\cdot,\nu) \,:\,D(\real^d;[0,1]) \mapsto \Theta$  is a
 Lipschitz-continuous operator
 in the ${\bf L}_{\infty}$ norm (where $D(\real^d;[0,1])$ is the Skorohod
space of functions);
\item[(ii)] 
For each $F \in D(\real^d;[0,1])$, the function 
$\theta(F,\cdot)$ is supported on $\cV$ and is a continuous
function; 
\item[(iii)] 
For any empirical distribution $\PP_N$, random variable
$\theta(\PP_N,\nu_N)$ induced
by random variable
$\nu_N$ satisfies (\ref{epsdelta_def}) with parameters
$(\varepsilon_N,\delta_N)$
\end{itemize}
\end{assumption}

Assumption \ref{operators}(i) focuses on data statistics which
continuously depend on the underlying data
distribution to guarantee that small changes
in that distribution leads to proportionally smaller
changes in the value of the output statistic. 
This, in turn, guarantees that for a fixed 
realization of the noise element $\nu$ convergence
of the empirical distribution leads to the 
convergence of the corresponding 
randomized statistic. Assumption \ref{operators}(ii) ensures that for each empirical
distribution of the data, the corresponding
randomized statistic is a continuous random
variable with respect to the distribution of the
induced noise. Therefore, the convergence
of the distribution of the random noise would
also lead to the convergence of random
variables induced by the randomized statistics. Assumptions \ref{operators}(i) and \ref{operators}(ii)  jointly ensure that the underlying statistics are not inherently difficult to deal with due to lack of continuity properties. Assumption \ref{operators}(iii) is the DP
guarantee. Notably, we allow the privacy parameters $(\varepsilon_N, \delta_N)$ to depend on the sample size $N$, consistent with both the theoretical DP  in the computer science literature and its practical implementations.

Definition~\ref{regularity} builds on  Assumption \ref{operators} and defined DP estimators that possess further desirable properties from the regularity perspective. Throughout, we consider DP parameters \((\varepsilon_N, \delta_N)\) such that \(\varepsilon_N \leq \bar{\varepsilon}\) and \(\delta_N \leq \bar{\delta}\) for all \(N\), where \(\bar{\varepsilon}\) and \(\bar{\delta}\) are fixed universal constants. This restriction is mild and aligns with standard practice.

\begin{definition}\label{regularity}
For a given sequence $\{(\varepsilon_N, \delta_N)\}$,  we say that an $(\varepsilon_N,\delta_N)$-differentially private estimator 
$\theta(\cdot,\cdot): {\mathcal Z}^N \times \cV \to \Theta$ satisfying Assumption \ref{operators}
is \textit{regular}  for the parameter of interest $\theta_0$ if the following conditions hold: 
\begin{enumerate}
\item[(i)] $\theta(\PP_N,\nu_N)$ is an ${\bf L}_1$-projection on $\Theta$ of a continuous random variable with respect to the Lebesgue measure;   
\item[(ii)]	For a given data-generating process, the estimator in the absence of mechanism noise (denoted as  $\theta(\PP_N, 0)$)  satisfies
    \begin{equation}\label{rate}
        \theta(\mathbb{P}_N, 0) \stackrel{p}{\longrightarrow} \theta_0,
    \end{equation}
    i.e., the mechanism noise-free estimator $\theta(\cdot, 0)$ is consistent for $\theta_0$.
\item[(iii)] $\theta(\PP_N,\nu_N)$ has a weak limit if the sequence $(\varepsilon_N,\delta_N)$ is convergent.
\end{enumerate}	
\end{definition}

Condition (i) ensures that the output of \(\theta(\cdot,\cdot)\) lies within the parameter space \(\Theta\). In cases where the randomization mechanism might otherwise push the estimator outside \(\Theta\), it is instead projected onto the boundary. Moreover, the requirement that the pre-projection distribution has a well-defined Lebesgue density ensures that the projection is well-defined and stable. Condition (ii) ensures that the noise-free estimator $\theta(\cdot, 0)$ is consistent, providing a reliable starting point, though it does not of course satisfy DP without noise. Finally, Condition (iii) imposes a basic regularity requirement and ensures the DP mechanism behaves sensibly in the large-sample limit and does not introduce pathological behavior.

\subsection{Example: Weighted sample mean of a random variable with bounded support}
\label{ex:samplemean2}

To illustrate the construction of randomized statistics that satisfy both Assumption~\ref{operators} and Definition~\ref{regularity}, as well as the subsequent steps of privacy-aware identification, we focus on estimating the weighted mean of a scalar random variable. The construction is presented from two complementary perspectives: that of the data curator, who ensures differential privacy, and the researcher, who employs the released statistics for inference. This example highlights features and challenges familiar to econometricians, particularly those arising in estimation procedures that involve tuning parameters and weighting schemes. 

Suppose \(X\) and \(W\) are independent one-dimensional random variables supported on \([0,1]\), and let the parameter of interest be the weighted mean \(\EE[X/W]\), which is assumed to belong to a known $[0,M]$ and  whose estimation is based on a sample of i.i.d.\ observations \(\{(x_i, w_i)\}_{i=1}^N\).

We consider four widely used Computer Science  mechanisms for constructing DP statistics \(\theta(\PP_N, \nu_N)\). A key challenge is that the ratio \(X/W\) can be unbounded, while most DP mechanisms require a bounded range. To ensure compatibility with these mechanisms, a pre-processing step is required. This step, performed by the data curator, involves truncating observations where \(w_i\) is close to zero. Specifically, for a sequence of thresholds \(\omega_N\), only observations with \(w_i \geq \omega_N\) are retained. The threshold \(\omega_N\) serves as a tuning parameter controlling the trade-off between bias and privacy.

Let \(F^x_N\) and \(F^w_N\) denote the empirical distributions of \(\{x_i\}\) and \(\{w_i\}\), respectively, and assume they converge as \(N \to \infty\). Define the effective sample size as \(n_N = N (1 - F^w_N(\omega_N))\), representing the expected number of retained observations after truncation. We normalize our estimators using \(n_N\).\footnote{Our results extend directly to using the realized sample size \(\sum_{i=1}^N \mathbf{1}(w_i \geq \omega_N)\), though this introduces additional technicalities without altering the main conclusions.} The noise-free estimator is $\theta(\PP_N, 0) = \frac{1}{n_N} \sum_{i=1}^N \frac{x_i}{w_i} \mathbf{1}\{w_i \geq \omega_N\}$. To comply with condition (ii) in Definition \ref{regularity}, we can suppose that $n_N \to \infty$, $\omega_N \to 0$. 

\smallskip 

\noindent {\bf 1. Laplace Mechanism.} This mechanism adds independent Laplace noise 
\( \nu_N \sim \text{Lap}\left(0, 1/(\varepsilon_N n_N \omega_N)\right) \) 
to the truncated sample mean:
\[
\theta(\PP_N, \nu_N) = \frac{1}{n_N} \sum_{i=1}^N \frac{x_i}{w_i} \mathbf{1}\{w_i \geq \omega_N\} + \nu_N,
\]
followed by projection onto \([0,M]\), consistent with the parameter space assumed for \(\EE[X/W]\). The first term represents the mean computed over the truncated sample, normalized by the effective sample size \(n_N\). Standard arguments (see \cite{dwork:2004}) verify that this mechanism satisfies \((\varepsilon_N, 0)\)-differential privacy.

\smallskip 

\noindent {\bf 2. Gaussian Mechanism.} This mechanism adds noise 
\( \nu_N \sim \mathcal{N}\left(0, 1/(\varepsilon_N^2 n_N^{2 - 2\gamma} \omega_N^2)\right) \) 
with \( 0 \leq \gamma \leq \tfrac{1}{2} \), and outputs
\[
\theta(\PP_N, \nu_N) = \frac{1}{n_N} \sum_{i=1}^N \frac{x_i}{w_i} \mathbf{1}\{w_i \geq \omega_N\} + \nu_N,
\]
followed by projection onto \([0,M]\). According to standard results (see \cite{dwork:2006}), this mechanism satisfies \((\varepsilon_N, \delta_N)\)-DP, where \( \delta_N = \Phi(-n_N^{\gamma} + 0.5 \varepsilon_N) \), and \(\Phi(\cdot)\) denotes the standard normal c.d.f.. Unlike the Laplace mechanism, the Gaussian mechanism does not achieve pure \((\varepsilon_N, 0)\)-DP. This is because the log-likelihood ratio of two Gaussians with equal variance but different means is unbounded in the mean difference. Therefore, it can be bounded within a fixed range only with a given probability.

\smallskip 

\noindent 
{\bf 3. Exponential mechanism} is non-additive unlike Laplace and Gaussian mechanisms above. It  
defines a sampling distribution
$$
p(z;\PP_N) \propto \exp\left(
-\frac{1}{2 }\epsilon_N^2 \omega_N^2
n_N^{2-2\gamma}
\left(
z-\frac{1}{n_N}\sum^N_{i=1}x_i/w_i
{\bf 1}\{w_i \geq \omega_N\}\right)^2
\right)
$$
and draws \(\theta(\PP_N, \nu_N) \sim p(\cdot; \PP_N)\), followed by projection onto \([0, M]\). We can apply the argument similar to that in the analysis of the Gaussian mechanism
to verify
that the output of 
exponential mechanism is
$(\epsilon,\Phi\left(-n_N^{\gamma}+.5\epsilon_N\right))$-DP.

\smallskip 

\noindent {\bf 4. ``Subsample and aggregate" mechanism.} This non-additive mechanism, introduced in \cite{nissim:07}, draws \(K_N\) independent subsamples \(S_k\) from the data and computes
\[
\theta_k = \frac{1}{|S_k|} \sum_{i \in S_k} \frac{x_i}{w_i} \mathbf{1}\{w_i \geq \omega_N\}, \quad k = 1, \ldots, K_N.
\]
The key insight in \cite{nissim:07} is that empirical statistics over the subsample estimates \(\{\theta_k\}\) are more robust to changes in individual observations.

Assuming e.g. equal-sized subsamples with \(n_N = K_N |S_k|\), one can take a robust summary—e.g., the median \(M(\theta_1, \ldots, \theta_{K_N})\)—and release a privatized version:
\[
\theta(\PP_N, \nu_N) = M(\theta_1, \ldots, \theta_{K_N}) + \nu_N,
\]
where \(\nu_N \sim \text{Lap}\left(0, 1/(\varepsilon_N^2 K_N \omega_N \max_k |S_k|)\right)\). Because the median is less sensitive to individual data points, this mechanism can achieve \((\varepsilon_N, 0)\)-differential privacy with potentially smaller noise than fully additive methods.

To summarize, \textit{from the 
perspective of a data curator}, as soon as truncation is conducted in the pre-processing stage, all mechanisms outlined here  behave identically to those in case
of an unweighted mean of a random
variable with a bounded range. 

From the \textit{researcher's perspective}, the situation is more complex. A researcher querying \(\mathbb{E}[X/W]\) typically knows the support of \(X\) and \(W\) (assumed public) and that truncation was applied, but may lack details about the truncation threshold \(\omega_N\), the empirical distributions \(F^x_N\) and \(F^w_N\), or the effective sample size \(n_N = N(1 - F^w_N(\omega_N))\). Without this information, the researcher must account for different possible statistical behaviors of the output \(\theta(\PP_N, \nu_N)\), which depend on the limiting behavior of the term \(\varepsilon_N n_N \omega_N\). These behaviors can be different even maintaining that $n_N \to \infty$, $\omega_N \to 0$ and can be loosely  classified into three regimes:

\begin{itemize}
    \item[\textbf{Regime 1:}] \(\varepsilon_N n_N \omega_N \to 0\) as \(N \to \infty\).  
    In this regime, the effective sample size after truncation is too small relative to the privacy parameter, causing the noise variance in all mechanisms to grow unbounded. Due to projection onto \([0, M]\), the output \(\theta(\PP_N, \nu_N)\) converges in distribution to a Bernoulli random variable taking values \(0\) or \(M\) with equal probability (\(1/2\)). This renders the estimate uninformative about \(\mathbb{E}[X/W]\).

    \item[\textbf{Regime 2:}] \(\varepsilon_N n_N \omega_N \to \infty\).  
    Here, the effective sample size is sufficiently large, and the noise variance in the Laplace, Gaussian, and Exponential mechanisms converges to zero. For the Subsample-and-Aggregate mechanism, a slightly stronger condition, \(\varepsilon_N^2 n_N \omega_N \to \infty\), ensures its noise variance also vanishes. Additionally, as \(\omega_N \to 0\), the truncated mean \(\frac{1}{n_N} \sum_{i=1}^N \frac{x_i}{w_i} \mathbf{1}\{w_i \geq \omega_N\}\) converges in probability to \(\mathbb{E}[X/W]\). Thus, the outputs of the Laplace, Gaussian, and Exponential mechanisms consistently estimate the target parameter.

    \item[\textbf{Regime 3:}] \(\varepsilon_N n_N \omega_N \to c\), where \(c\) is a finite positive constant.  
    This intermediate case occurs when the effective sample size and privacy parameter balance out, leading to noise variances in the Laplace, Gaussian, and Exponential mechanisms converging to a constant. The output \(\theta(\PP_N, \nu_N)\) is distributed across the parameter space \([0, M]\), potentially with positive probability at the boundaries. For the Subsample-and-Aggregate mechanism, this regime applies unless \(\varepsilon_N \to 0\), in which case it reverts to Regime 1.
\end{itemize}

If \(\varepsilon_N n_N \omega_N\) lacks a limit, the researcher must consider all possible partial limits along converging subsequences, potentially encountering a mixture of these regimes. 

To summarize, the researcher faces uncertainty about the statistical properties of \(\theta(\PP_N, \nu_N)\), as the output could reflect a random set of behaviors driven by the mechanism and regime, even with known DP constraints.

This example underscores the need for a researcher to consider any given DP output as a realization of just some selection from a random set comprised of many possible DP statistics.

\subsection{Direct privacy-aware identification approach (no additional data curator involvement)}
\label{sec:directID}

To study the notion of identification, we now define a random set of DP estimators, reflecting the researcher's perspective. This set is shaped by the researcher's knowledge or inferences about the DP constraint sequences \((\varepsilon_N, \delta_N)\) chosen by the data curator. We encapsulate this knowledge in a fixed collection of sequences, denoted \(\mathcal{E}\), which may include a single sequence or multiple sequences (e.g., when DP parameters are communicated as upper bounds, satisfying the DP condition ~\eqref{epsdelta_def} with a strict inequality). As illustrated in our example in Section \ref{ex:samplemean2}, different asymptotic regimes arise from the interplay between the asymptotic behavior of \((\varepsilon_N, \delta_N)\) and the truncation process. Consequently, the choice of \(\mathcal{E}\) influences the random set considered by the researcher, as certain \(\mathcal{E}\) may exclude specific asymptotic regimes in a model.

Following \cite{molchanov2005}, we use the concept of \textit{measurable selection} to define the set of all regular DP estimators that the researcher considers possible, given the known collection \(\mathcal{E}\) and any additional information communicated by the data curator (e.g., partial details of the implemented mechanism). This comprehensive perspective is essential for analyzing identification, as it accounts for all admissible estimators under the given constraints, rather than focusing on the properties of specific DP procedures.

\begin{definition}\label{def:random:set}
Let $\rT^*_{N, \mathcal{E}}$ denote the set of all random variables $\theta(\PP_N, \nu_N): \mathcal{Z}^N \times \mathcal{V} \rightarrow \Theta$ and correspond to sequences $(\varepsilon_N, \delta_N) \in \mathcal{E}$.  In case other information about the DP mechanism is available, they have to be admissible under that information. 

Define $\rT_{N, \mathcal{E}}$ as the completion of $\rT^*_{N, \mathcal{E}} \cap \mathbf{L}_1(\PP)$ with respect to the $\mathbf{L}_1(\PP)$-norm, where 
 $\mathbf{L}_1(\PP)$ denotes the space of measurable functions from $\mathcal{Z}^N \times \mathcal{V}$ into $\mathbb{R}^p$ that are integrable with respect to the product measure induced by the distribution on $\mathcal{Z}$ and random $\nu_N$.\footnote{By the properties of $\theta(\cdot, \cdot)$ in Definition~\ref{regularity} and the compactness of $\Theta$, all elements in $\rT^*_{N, \mathcal{E}}$ (and, hence in $\rT_{N, \mathcal{E}}$. are bounded in the $\mathbf{L}_1$-norm.}

We refer to $\rT_{N, \mathcal{E}}$ as the \textit{set of regular DP estimators for $\theta_0$ under $\mathcal{E}$}.
\end{definition}

 $\rT_{N, \mathcal{E}}$ is a compact random set in the sense of Definitions 1.30 in \cite{molchanov2005}, as shown in Lemma \ref{lemma:TNconvexcompact}. 
\begin{lemma}
\label{lemma:TNconvexcompact}	
$\rT_{N, \mathcal{E}}$  is a compact random set. If  $\mathcal{E}$ is a join-semilattice in the coordinate-wise partial order for $(\varepsilon_N,\delta_N)$, then $\rT_{N, \mathcal{E}}$  is also a  convex random set. 
\end{lemma}

We are now ready to introduce the concept of \textit{identification} for DP estimators. As conveyed earlier, we approach this notion from the perspective of a limit which ultimately summarizes what a researcher could learn from an observation from a random set $\rT_{N, \mathcal{E}}$ given access to an infinitely large sample. This concept must be grounded in some well-defined notion of the limit of random sets and, at first glance, it may seem natural to use the probability limit  to formalize this idea with the hope of relating identifiability of the parameter to the event that the random sets $\rT_{N, \mathcal{E}}$ intersect any neighborhood of $\theta_0$ with probability approaching $1$. However, we find the probability limit  too restrictive for typical DP estimators as The presence of the noise term $\nu_N$ whose distribution does not degenerate as $N \to \infty$ would often prevent the estimator from converging to a simple probability limit  necessitating an alternative limit notion.

A natural notion of limit for our purposes is the weak limit, denoted by $\stackrel{W}{\rightarrow}$ throughout the paper. In fact, we have already   indicated our focus on weak limits in condition (iii) of Definition \ref{regularity}. Lemma \ref{lemma:weaklimit} establishes that weak convergence is the strongest reasonable convergence concept to consider, unless the weak limits of regular differentially private estimators are constant with probability 1.

\begin{lemma} 
\label{lemma:weaklimit}	
Suppose that $\theta(\PP_N,\nu_N) \stackrel{W}{\rightarrow} \tau$ as $N \rightarrow \infty.$ Then if $\tau$ is not constant with  probability 1, then there exists $\bar{\kappa}>0$ and
$\gamma>0$ such 
that for all $\kappa \leq \bar{\kappa}$
$$
\limsup\limits_{N \rightarrow \infty}\PP\left(
|\theta(\PP_N,\nu_N)-\tau|>\kappa
\right)>\gamma.
$$
\end{lemma}

Our next result, Theorem \ref{convergence}, justifies the use of weak convergence and establishes the weak convergence of the convex compact random set $\rT_{N,\mathcal{E}}$ under mild conditions on $\mathcal{E}$.
\begin{theorem}\label{convergence}
Let $\mathcal{E}$ be a join-semilattice consisting solely of convergent sequences $(\varepsilon_N, \delta_N)$. Then:
\begin{itemize}
\item[(i)] The random set $\rT_{N,\mathcal{E}}$ converges weakly to a limit denoted by $\rbT_{\mathcal{E}}$.

\item[(ii)]  $\rbT_{\mathcal{E}}$ is a convex random set which is the closure of all weak limits of estimators in the corresponding random set $\rT_{N,\mathcal{E}}$.

\end{itemize}
\end{theorem}
The convergence result in Theorem \ref{convergence} characterizes the limiting behavior of statistical experiments. We therefore define identifiability through properties of the random set $\rbT_{\mathcal{E}}$, which captures what can be inferred from an infinitely large sample. If $\rbT_{\mathcal{E}}$ degenerates at $\theta_0$, them the experiments asymptotically reveal the true parameter. In practice, however, the limit set may exclude $\theta_0$ (``biased'') or include it without degenerating at $\theta_0$ (``inconsistent''). 

Which properties of random set $\rbT_{\mathcal{E}}$ are most suitable for our purposes? Prior econometric work involving random sets (e.g., \cite{BeresteanuMolinari08}, \cite{beresteanu2012}, among many subsequent work that followed these papers) measures information content of random sets via the {selection expectation}. In fact, our earlier work \cite{KNY:18} adopted this selection expectation approach to study privacy under data combination, where access to all datasets made it natural. In the DP setting, however, the privacy budget across multiple statistics renders this approach infeasible, as we discuss in more detail at teh end of this section.  Instead, to characterize $\rT_{N,\mathcal{E}}$ and $\rbT_{\mathcal{E}}$, we adopt the {containment functional} from \cite{molchanov2005}:

\begin{definition}[Molchanov (2005), Definition 1.32]
\label{coverage}
For a compact $K \subset \Theta$, the containment functional of random set $\rbX$ is  $C_{\rbX}(K)=\PP\left(\rbX \subset K\right)$.
\end{definition}

By Proposition 1.1.33 in \cite{molchanov2005}, this functional fully characterizes the distribution of a compact random set. By Theorem 1.7.8 for a convex compact random set  it is enough to only consider convex polytopes  as ``test sets'' $K$ in the definition above. It also preserves weak convergence:  

\begin{theorem}\label{coverage:convergence}
Under Theorem \ref{convergence}, for any convex polytope $K \subset \Theta$,  $C_{\rT_{N,\mathcal{E}}}(K) \to C_{\rbT_{\mathcal{E}}}(K)$,  $N \to \infty$.
\end{theorem}

This result (a corollary of Theorem 1.6.5 in \cite{molchanov2005}) ensures that convergence of random sets $\rT_{N,\mathcal{E}}$ is equivalently captured by pointwise convergence of their containment functionals. Thus, analyzing the limit of $\rbT_{\mathcal{E}}$ reduces to analyzing its containment functional on convex polytopes in $\Theta$. 

We illustrate $\rbT_{\mathcal{E}}$ and its containement functional in the example below. 

\smallskip 

\textbf{EXAMPLE in Section \ref{ex:samplemean2} (continued).} \textit{In Example in Section \ref{ex:samplemean2}, we estimated the weighted mean $\theta_0 = \EE[X/W]$ under differential privacy in an asymptotic regime where finite sample distributions $F^x_N(\cdot)$ and $F^w_N(\cdot)$ converge as $N \rightarrow \infty$. Observations with $w_i < \omega_N$ were discarded using an adaptive truncation threshold $\omega_N$ to ensure finite noise addition for privacy guarantees.} 

\textit{Suppose what available to a researcher is $\rT^*_{N, \mathcal{E}} = \text{co}\{\theta^{(1)}(\PP_N, \nu_N), \theta^{(2)}(\PP_N, \nu_N)\}$, where $\theta^{(1)}(\PP_N, \nu_N)$ and $\theta^{(2)}(\PP_N, \nu_N)$ denote outputs in regimes 1 and 2, respectively, obtained e.g. by the Laplace DP mechanism and are $(\epsilon_N, 0)$-differentially private.  Just like in Section \ref{ex:samplemean2}, a researcher can know that $n_N \to \infty$ and $\omega_N \to 0$.}

\textit{In this case, of course, $\rT_{N, \mathcal{E}} \equiv \rT^*_{N, \mathcal{E}}$. As discussed in Section  \ref{ex:samplemean2}, $\theta^{(1)}(\PP_N, \nu_N)$ converges in distribution to a Bernoulli random variable taking values 0 or $M$ with
equal probability ($1/2$), and $\theta^{(2)}(\PP_N, \nu_N)$ converges in probability to $\theta_0 \in [0,M]$. Thus:
\[
(\theta^{(1)}(\PP_N, \nu_N), \theta^{(2)}(\PP_N, \nu_N)) \stackrel{W}{\longrightarrow} (B_{\{0,M\}}(0.5), \theta_0).
\]
The limiting random set is
\[
\rbT_{\mathcal{E}} = \begin{cases} 
[0, \theta_0], & \text{with probability } 1/2, \\
[\theta_0, M], & \text{with probability } 1/2, 
\end{cases}
\]
and its containment functional is
\[
C_{\rbT_{\mathcal{E}}}(K) = \frac{1}{2} \mathbf{1}\{[0, \theta_0] \subset K\} + \frac{1}{2} \mathbf{1}\{[\theta_0, M] \subset K\}. \blacksquare
\]}
This example illustrates that regular differentially private estimators converge weakly to a limiting random set $\rbT_{\mathcal{E}}$ whose distribution is non-degenerate.

Generally, we shall view the random set $\rbT_{\mathcal{E}}$ obtained as the weak limit of $\rT_{N,\mathcal{E}}$ as a \textit{pseudo-identified set}.  Unlike standard econometrics, where identified sets (or in some misspecified models  pseudo-identified sets that don't contain the true parameter) are deterministic,  the random pseudo-identified set is more suitable in the DP setting due to combined sampling and mechanism noise which often cannot be separated. Related work, such as \cite{KitagawaLowerUpper}, has also considered random identified sets, though there the randomness stems from posterior uncertainty. 

We now define identifiability under differential privacy. 

\begin{definition}[Identifiability of parameters under DP]\label{identification}
Let $\mathcal{E}$ be a join-semilattice that consists of converging sequences $\{(\varepsilon_N, \delta_N)\}$. Parameter $\theta_0$  is identified in the regular $(\varepsilon_N,\delta_N)$-DP framework, where sequences  $\{(\varepsilon_N,\delta_N)\}$ belong to $\mathcal{E}$, if and only if for any $\alpha \in (0,1)$ and any convex polytope $K \ni \theta_0,$ we have $C_{\rbT_{\mathcal{E}}}(K) \geq 1 - \alpha$.
\end{definition} 
Theorem \ref{th:iff} links identifiability to the convergence of random sets to a singleton, akin to consistency for random variables.
\begin{theorem}\label{th:iff}
Suppose the conditions of Theorem \ref{convergence} hold. For any sequence $\{(\varepsilon_N,\delta_N)\}$ from $\mathcal{E}$ it holds that any regular   $(\varepsilon_N,\delta_N)$-DP estimator $\theta(\PP_N,\nu_N) $ satisfies $\theta(\PP_N, \nu_N) \stackrel{p}{\rightarrow} \theta_0$ 
if and only if, for any $\alpha \in (0,1)$ and any convex polytope $K \ni \theta_0$ we have
$C_{\rbT_{\mathcal{E}}}(K)  \geq 1-\alpha$ and, thus, parameter $\theta_0$ is identifiable even under differential privacy.  
\end{theorem}

Based on the same principles we can characterize the case of non-identifiability.

\begin{definition}[Non-identifiability under DP]\label{nonidentification}
Let $\mathcal{E}$ be a join-semilattice and consist of converging sequences of $\{(\varepsilon_N,\delta_N)\}$.  Parameter $\theta_0$ is non-identified in the regular $(\varepsilon_N, \delta_N)$-DP framework, where the sequences $\{(\varepsilon_N,\delta_N)\}$ belong to $\mathcal{E}$, if there exists $\beta \in (0,1)$ and a convex polytope $K_{\beta} \ni \theta_0$ such that $C_{\rbT_{\mathcal{E}}}(K_{\beta}) \leq 1 - \beta$.
\end{definition}

Non-identifiability means that the limiting random set is non-degenerate, so $\theta_0$ cannot be recovered as a ``mass point'' of the containment functional. This differs from traditional partial identification, which constructs deterministic sets containing $\theta_0$. A non-degenerate containment functional in DP makes this infeasible. As the containment functional may  be difficult to work with in practice, in one of subsequent sections we develop a more tractable approach to identification and to the construction of Bayesian credible regions.

\paragraph*{Privacy budget and the inapplicability of expectation-based notions for randoms sets}

Having introduced identifiability under DP, we return to the role of expectation-based notions of random sets, such as the selection (Aumann) expectation. We also discuss Vorob’ev’s expectation. 
The selection expectation is particular is widely used when identifiability is linked to random sets (see e.g.  \cite{BeresteanuMolinari08}, \cite{beresteanu2012} and subsequent literature inspired by these papers). While such expectation-based notions are natural in non-DP settings,  they are unsuitable under DP due to privacy budget constraints as, unlike in classical statistics, privacy budget considerations imply  that a DP mechanism cannot be freely replicated or re-run to approximate expectations by repeated averaging. This is something we have already touched upon briefly in Section \ref{sec:DPreview} when discussing multiple queries and privacy budget. 

Let us elaborate further on why in a DP setting an expectation-based approach fails to be representative. This limitation arises because with a given  privacy budget differential privacy is guaranteed at the dataset level, not for individual statistics in isolation. By the composition property, the privacy parameters of multiple statistics aggregate across queries. For example, two $(\varepsilon/2,0)$-DP statistics together form at most a $(\varepsilon,0)$-DP release. More generally, computing $K$ statistics under a fixed $(\varepsilon,0)$ budget may require each to satisfy $(\varepsilon_k,0)$-DP with $\sum_{k=1}^K \varepsilon_k =\varepsilon$ which demands greater noise per statistic. A natural way to preserve the privacy budget is for the curator to generate the random seed $\nu_N$ once and keep it secret. The statistic $\theta(\PP_N,\nu_N)$ is then fixed for any query of the same form, ensuring that repeated evaluation yields the same output. While this guarantees DP for the entire collection of statistics based on $\nu_N$, it also means that expectations such as $\EE_{\nu_N}[\theta(\PP_N,\nu_N)]$ are fundamentally unobservable. Consequently, notions such as the selection expectation or Vorob’ev’s expectation are misleading in the DP context as they rely on averaging across realizations that privacy guarantees explicitly prohibit.  This underscores the necessity of using the containment functional, rather than expectation-based notions, to characterize $\rbT_{\mathcal{E}}$ under DP.

To sum up, the containment functional is the most appropriate characterization of the limiting random set $\rbT_{\mathcal{E}}$.

\subsection[Smoothness property]{Important case: Smooth DP estimators}
\label{sec:smoothness}

While our previous discussion considered a general, potentially non-separable form of regular DP estimators, all practical DP mechanisms in use exhibit an approximately separable structure (to the best of our knowledge). As we demonstrate in this section, this property greatly simplifies the analysis of identification. We refer to this approximate separability as \textit{smoothness} of DP estimators and show how it enables the explicit construction of the limiting random set $\rbT_{\mathcal{E}}$, facilitating the study of identifiability.

\begin{definition}\label{smoothness}
A regular differentially private estimator \(\theta(\PP_N, \nu_N) \in \rT_{N,\mathcal{E}}\) is \textit{smooth} if it can be expressed as
\[
\theta(\PP_N, \nu_N) = \psi(\PP_N) + a(\nu_N) + \Delta_N,
\]
where \(\EE[\|\Delta_N\|_{\infty}^2] \to 0\) as \(N \to \infty\), and
\begin{itemize}
    \item[(i)] \(\psi(\cdot)\) is a Lipschitz functional such that, for any sequence of distributions \(\mathbb{F}_N \stackrel{W}{\to} F\) it follows that \(\psi(\mathbb{F}_N) \stackrel{W}{\to} \psi(F)\).
    \item[(ii)] \(a(\cdot)\) is a continuous, bounded measurable functions, with universal continuous upper and lower envelopes \(A_{*}(\cdot) \leq a(\cdot) \leq A^*(\cdot)\).
\end{itemize}
\end{definition}

The smoothness property characterizes the approximate separability of DP estimators 
into a data-dependent component $\psi(\PP_N)$ and a mechanism noise component $a(\nu_N)$, 
with a negligible residual term $\Delta_N$. 
This structure arises naturally in many DP mechanisms (to the best of our knowledge, all most common practical DP mechanisms satisfy this property), 
where $\psi(\PP_N)$ often corresponds to the noise-free statistic of interest, 
$\theta(\PP_N, 0)$, and the smoothness condition formalizes the addition of 
calibrated noise required for privacy while preserving desirable statistical properties. 
Crucially, smoothness enables a tractable analysis of the weak limit of $\rT_{N,\mathcal{E}}$, 
which is central to assessing identifiability under DP, 
as established in Theorem~\ref{convergence} and Definition~\ref{identification}.

The structure of the randomized statistic
$\theta(\PP_N,\nu_N)$ in Definition \ref{smoothness}
is closely related to the {\it influence function
representation.} Randomized statistics which are 
derived from non-private estimators for which 
such an influence function representation exists
will have representation conforming to that 
in Definition \ref{smoothness}.

Theorem \ref{representation:smooth} derives an explicit representation of the limiting random set \(\rbT_{\mathcal{E}}\), which simplifies the study of its containment functional \(C_{\rbT_{\mathcal{E}}}(K) = \PP(\rbT_{\mathcal{E}} \subset K)\).

\begin{theorem}\label{representation:smooth}
Consider a class of smooth DP estimators \(\theta(\PP_N, \nu_N) \in \rT_{N,\mathcal{E}}\) as in Definition~\ref{smoothness} for any sequence \(\{(\varepsilon_N, \delta_N)\} \in \mathcal{E}\). Let \(\Psi\) denote the convex hull of the weak limits of \(\psi(\PP_N)\), and let \(\mathcal{A}\) denote the convex hull of weak limits of \(A_*(\nu_N)\) and \(A^*(\nu_N)\). Then, the limiting random set \(\rbT_{\mathcal{E}}\), as defined in Theorem~\ref{convergence}, is the Minkowski sum:
\[
\rbT_{\mathcal{E}} = \Psi \oplus \mathcal{A}.
\]
\end{theorem}

Theorem~\ref{representation:smooth} implies that the limiting random set $\rbT_{\mathcal{E}}$ 
can be constructed by separately analyzing the weak limits of the data-dependent component 
$\psi(\PP_N)$ and the privacy noise component $a(\nu_N)$. 
This decomposition is particularly useful for identification, as it allows researchers to 
characterize $\rbT_{\mathcal{E}}$ without having to analyze each estimator individually. 
Instead, the containment functional $C_{\rbT_{\mathcal{E}}}(K)$ can be studied through the 
convex hulls $\Psi$ and $\mathcal{A}$, which capture the asymptotic behavior of the 
data-dependent and noise components, respectively. 

For identifiability (Definition~\ref{identification}), 
$\rbT_{\mathcal{E}}$ must degenerate to a singleton at $\theta_0$, 
which requires both $\Psi$ and $\mathcal{A}$ to collapse appropriately under the constraints 
imposed by $\mathcal{E}$. 
Typically, identifiability is achieved when $\Psi$ collapses to $\{\theta_0\}$ 
(as is intuitive when $\psi(\PP_N) = \theta(\PP_N, 0)$) 
and $\mathcal{A}$ collapses to $\{0\}$ 
(as is intuitive when the mechanism noise variance vanishes, causing the noise to become 
increasingly concentrated near zero).

\paragraph*{Smoothness in common DP practical mechanisms}

To the best of our knowledge all widely used DP mechanisms naturally satisfy the smoothness condition, which simplifies the application of Theorem~\ref{representation:smooth} in practical settings. Below, we discuss how common DP mechanisms outlines in the example in Section ~\ref{ex:samplemean2}   align with the smoothness property and contribute to the identification framework.

\vskip 0.05in

 \noindent \textbf{1. Laplace and Gaussian mechanisms}: In the Laplace mechanism, differential privacy is achieved by adding double-exponential noise to a non-private statistic, while the Gaussian mechanism uses normal noise. For both, the estimator can be written as \(\theta(\PP_N, \nu_N) = \theta(\PP_N, 0) + a(\nu_N)\), with \(\Delta_N \equiv 0\), satisfying Definition~\ref{smoothness} trivially. The noise component \(a(\nu_N)\) has well-defined envelopes (e.g., the support of the Laplace or Gaussian distribution), and the data-dependent component \(\theta(\PP_N, 0)\) is typically a Lipschitz functional, such as the sample mean or median. In the context of identification, the weak limit of \(\theta(\PP_N, 0)\) determines whether \(\Psi\) collapses to \(\theta_0\) (it does so under conditions in Definition \ref{regularity}), while the noise envelopes determine the spread of \(\mathcal{A}\). For example, in the setting of example in Section ~\ref{ex:samplemean2}, the Laplace mechanism’s noise may contribute to the non-degenerate limiting set \(\rbT_{\mathcal{E}}\) through a non-degenerate $\mathcal{A}$, leading to non-identifiability. If the mechanism noise variance vanishes, then $\mathcal{A}=\{0\}$ and we have identifiability.

\vskip 0.05in

\noindent  \textbf{2. Subsample and Aggregate Mechanism}: Proposed by \cite{nissim:07}, this mechanism splits the data into \(K\) independent subsamples, computes a non-private statistic \(\widehat{\theta}_k\) on each subsample \(k\), and aggregates them using an aggregation function \(f_K(\cdot)\), such as the median, with calibrated noise added for privacy.  This mechanism
is smooth according to our definition
with $\psi(\PP_N)=f_K(\widehat{\theta}_1,
\ldots,\widehat{\theta}_K)$ and an appropriately
scaled
additive noise $a(\nu_N).$ 
The weak limit of
$\psi(\PP_N)$ depends both on the convergence
of statistics $\widehat{\theta}_k$ and the 
asymptotic behavior of the aggregation function 
which also depends on the limit of the number
of subsamples $K.$ This structure, once again, allows the researcher to assess whether \(\rbT_{\mathcal{E}} = \Psi \oplus \mathcal{A}\) degenerates to \(\theta_0\), as required for identifiability.

\vskip 0.05in 

\noindent \textbf{3. Exponential Mechanism}: One prominent example of a non-separable mechanism for DP is the 
\textbf{exponential mechanism} introduced in \cite{mcsherry:07}. Applied to extremum estimators, it replaces the maximizer 
$\widehat{\theta}$ of the sample objective function $Q(\theta;\,\PP_N)$ with a draw from a 
\emph{quasi-posterior} distribution derived from $Q(\cdot;\,\PP_N).$ This approach is closely related to randomized estimators in \cite{CH:03}.  \cite{CH:03} consider cases where the population objective satisfies the information matrix equality. They construct an estimator by introducing a prior $\pi(\cdot)$ on $\theta$ and defining a quasi-likelihood 
$\exp(Q(\theta;\,\PP_N))$, so that $Q$ acts as a quasi-log-likelihood. The resulting quasi-posterior  $\propto \exp(Q(\theta;\,\PP_N)) \pi(\theta)$
yields a posterior mean that consistently estimates the population maximizer under mild regularity conditions, regardless of $\pi(\cdot).$ Moreover, the quasi-posterior variance consistently estimates the asymptotic variance of $\widehat{\theta}.$ A key advantage is that this method avoids direct maximization of potentially non-smooth or computationally difficult objective functions. \footnote{ Building on this, \cite{KN:16} address situations where $Q(\theta;\,\PP_N)$ is steep near its maximum, which can slow convergence of Markov chain sampling from the quasi-posterior. They propose rescaling the exponent to  $\exp(\lambda\,Q(\theta;\,\PP_N))$, where $\lambda$ is chosen to improve mixing. The posterior mean remains consistent for the population maximizer, and its asymptotic variance is estimated by rescaling the quasi-posterior variance with $\lambda.$}  

The exponential mechanism for DP considered in  \cite{mcsherry:07} is a
simple implementation of the idea in \cite{CH:03}: the estimator is a single draw from the 
quasi-posterior $\propto \exp(\lambda\,Q(\theta\,;\,\PP_N))\,\pi(\theta).$ The resulting
estimator turns out to be $\left(\lambda \Delta\,Q\,,\,0\right)$-differentially private, 
where
$$\Delta\,Q=\sup\limits_{\theta \in \Theta,\,\PP_N,\PP_N'}\left|Q(\theta\,;\,\PP_N')-Q(\theta\,;\,\PP_N) \right|$$
is the {\it global sensitivity} of the objective function $Q(\theta\,;\,\PP_N)$ evaluated
over all empirical distributions $\PP_N'$ that are different from $\PP_N$ in any one single support point if $\Delta\,Q$ is bounded.
If $\Delta\,Q$ is unbounded,
the estimator is $(\epsilon,\delta)$
-differentially private 
with vanishing $\delta$
for an appropriately calibrated
scaling constant $\lambda.$

\cite{CH:03} focus on the cases where $Q(\theta\,;\,\PP_N)$ is
stochastically equicontinuous and the quasi-posterior is asymptotically equivalent to 
$$\exp\left(-0.5\lambda (\theta-\widehat{\theta})'H \, (\theta-\widehat{\theta})+o_p(\|\theta-\widehat{\theta}\|^2)\right),
$$ 
where $H$ is the Hessian of the population objective function.
This means that a single draw from this quasi-posterior, corresponding to the 
exponential mechanism for differential privacy can be represented as
$\tilde{\theta}=\widehat{\theta}+\lambda\,\xi+o_p(1),
$ 
where $\xi$ is a multivariate normal random vector with mean zero and covariance matrix $H^{-1}.$
The extremum estimator $\widehat{\theta}$ only depends on the data distribution $\PP_N$ and is 
not affected by the noise.
Therefore, the exponential mechanism is smooth in the sense of Definition \ref{smoothness}.  

\subsection{Identifiability with the active collaboration for a data curator} 
\label{sec:datacurator}

As we demonstrate in the applications
in Section \ref{sec:appl}, the setting where
the limiting random set
$\rbT_{\mathcal{E}}$ is not a 
singleton is not an exception but
 may be a commonplace
scenario. In such settings, a smooth regular DP estimator $\theta(P_N, \nu_N)$ need not concentrate near the true parameter $\theta_0$, even as $N \to \infty$. More importantly, since $\rbT_{\mathcal{E}}$ itself is random, events such as $\{\rbT_{\mathcal{E}} \subset K \}$ for convex polytopes $K$ can occur with probability strictly between zero and one. This rules out direct use of standard tools for partial identification within a general DP framework as the parameter of interest is neither point-identified nor partially identified in the classical sense.

Current DP practices, such as those employed by the U.S. Census Bureau (2021), emphasize \textit{verifiable protocols} which are processes that take raw data as input and produce randomized statistics with traceable noise infusion to ensure DP guarantees. While this verifiability allows outsiders to retrace steps from raw data to output, it alone does not suffice for identifying the target parameter as substantial informational asymmetries may still remain in place. 

To address this, we propose a refinement that restores point identification by leveraging the structure of the limiting random set of DP statistics. Specifically, we introduce a decision functional that maps the random set $\rT_{N,\mathcal{E}}$  to a single selection $\theta(P_N, \nu_N)$ from this set.

\begin{definition}\label{def:decision:map}
Mapping $\tau_f$ from the elements of Fell topology on $\Theta$ into $\Theta$ indexed
by a $\alpha$-strongly convex function $f$ for some $\alpha>0$ on $\Theta$
such that for a fixed realization $\rT_{N,\mathcal{E}}(\omega)$
of $\rT_{N,\mathcal{E}}$
$$
\tau_f\left(\rT_{N,\mathcal{E}}(\omega)\right)=
\Argmin\limits_{z \in \rT_{N,\mathcal{E}}(\omega)}
f(z)
$$
is referred to
as the \textit{decision mapping} of the data curator.
\end{definition} 

This definition captures two essential aspects of ``transparent behavior'' for a data curator aiming to provide DP guarantees: (1) the curator must fully explore and understand the entire random set $\rT_{N,\mathcal{E}}$, which contains all regular differentially private estimators for the target parameter; and (2) the curator selects a point within this set using a principled approach based on the set's geometry, by minimizing the convex function $f(\cdot)$, which is publicly communicated.

This approach differs fundamentally from a ``transparent reproducible algorithm'' for DP (e.g., as in U.S. Census Bureau, 2021), which merely selects an arbitrary element of $\rT_{N,\mathcal{E}}$ without ensuring it holds a special position (e.g., the center of gravity). As follows from our results below, without knowledge of the selected output's relative position in $\rT_{N,\mathcal{E}}$, identification of $\theta_0$ is impossible.

\begin{lemma} \label{lemma:DCmeasselec}
The decision mapping $\tau_f(\rT_{N,\mathcal{E}})$ is a random variable.
\end{lemma}

We next examine how the convergence of the random sets $\rT_{N,\mathcal{E}}$ to the limit $\rbT_{\mathcal{E}}$ implies convergence of the decision mapping's realizations. Note that $\tau_f(\rT_{N,\mathcal{E}}) \leq t$ if and only if $\rT_{N,\mathcal{E}}$ intersects the convex compact set $\{ f(z) \leq t \}$. In other words,
\[
P(\tau_f(\rT_{N,\mathcal{E}}) \leq t) = P(\rT_{N,\mathcal{E}} \cap \{ f(z) \leq t \} \neq \emptyset) = 1 - C_{\rT_{N,\mathcal{E}}}(\{ f(z) > t \}).
\]
Combining this with Theorem  \ref{convergence} yields the following:

\begin{theorem} 
\label{th:DCweakconv}
The sequence of random variables $\tau_f(\rT_{N,\mathcal{E}})$ converges weakly as $N \to \infty$, with limit $\tau_f(\rbT_{\mathcal{E}})$.
\end{theorem}
(The proof is omitted, as it follows directly from the preceding results.)

A key question is whether the weak limit $\tau_f(\rbT_{\mathcal{E}})$ preserves DP with privacy parameters corresponding to the limits of sequences $\{(\varepsilon_N, \delta_N)\}$ in $\mathcal{E}$. Since each element of $\rbT_{\mathcal{E}}$ is differentially private with parameters determined by $\mathcal{E}$, we invoke the post-processing property of DP \citep[Proposition 2.1]{dwork2014algorithmic} to establish the following:

\begin{corollary}
\label{corollary:decisionmapping}
$\tau_f(\rbT_{\mathcal{E}})$ is differentially private with privacy parameters determined by the limits of sequences in $\mathcal{E}$.
\end{corollary}

The results of Theorem \ref{th:DCweakconv} and Corollary \ref{corollary:decisionmapping} are useful for identifiability if $f(\cdot)$ ensures that the limiting distribution of $\tau_f(\rT_{N,\mathcal{E}})$ is degenerate at $\theta_0$. Selecting an appropriate $f(\cdot)$ that is tailored to the structure of $\rbT_{\mathcal{E}}$ for the given problem is the critical task of the data curator to make the DP framework ``useful'' for estimating $\theta_0$.

 How can the data curator proceed?  The curator's distributional knowledge of $\rbT_{\mathcal{E}}$ means that $C_{\rbT_{\mathcal{E}}}(\cdot; \theta)$ is known for a generic parameter $\theta$ of the data-generating process.  Thus, a curator with full knowledge of $\rbT_{\mathcal{E}}$'s distribution can choose $f(\cdot)$ such that 
$P(\tau_f(\rbT_{\mathcal{E}}) = \theta \mid \theta) = 1$ 
( e.g., $\theta$ could be located relative to extreme points  realizations of $\rbT_{\mathcal{E}}$). 
Then,  for each realization of $\rT_{N,\mathcal{E}}$, the data curator geometrically locates $\tau_f(\rT_{N,\mathcal{E}})$ within it.
theorem \ref{th:DCconsistency} establishes that this decision mapping ensures the selected regular DP output $\tau_f(\rT_{N,\mathcal{E}})$ approximates $\theta_0$ with high probability.

\begin{theorem} 
\label{th:DCconsistency}
Suppose the data curator's decision rule satisfies $P(\tau_f(\rbT_{\mathcal{E}}) = \theta \mid \theta) = 1$ for $\theta \in \Theta$. Then $\tau_f(\rT_{N,\mathcal{E}}) \xrightarrow{p} \theta_0$.
\end{theorem}

To illustrate how a data curator can choose $f$, consider a simplified example with a closed-form limiting random set $\rbT_{\mathcal{E}}$ and a suitable decision function.

\vskip 0.05in 

\textbf{Example (continuation of example in Section \ref{ex:samplemean2})}. \textit{
Returning to the estimation of the weighted mean, where all elements of the limiting random set are $(\varepsilon_N, 0)$-DP, the limit takes the form
\[
\rbT_{\mathcal{E}}  = 
\begin{cases} 
[0, \theta_0] & \text{with probability } 1/2, \\ 
[\theta_0, M] & \text{with probability } 1/2.
\end{cases}
\]
This structure underscores our point: if the data curator fully explores the distribution of $\rbT_{\mathcal{E}}$,  they know the containment functional $C_{\rbT_{\mathcal{E}}}(K) = \frac{1}{2} \mathbf{1}\{ [0, \theta_0] \subset K \} + \frac{1}{2} \mathbf{1}\{ [\theta_0, M] \subset K \}$. By constructing a decision mapping that selects the extremal point not near the parameter space boundaries (0 or $M$), the curator can consistently estimate $\theta_0$. For instance, the data curator can choose to minimize the following 1-strongly convex function 
\[
f(z) = \left( z - \min_{z \in \rbT_{\mathcal{E}}} z \cdot \mathbf{1}\{ |\max_{z \in \rbT_{\mathcal{E}}} z - M| < h \} - \max_{z \in \rbT_{\mathcal{E}}} z \cdot \mathbf{1}\{ |\min_{z \in \rbT_{\mathcal{E}}} z| < h \} \right)^2
\]
for small $h > 0$. This attains its minimum at $\theta_0$ for any realization of $\rbT_{\mathcal{E}}$. Minimizing this $f$  over finite sample $\rT_{N,\mathcal{E}}$ yields a consistent, regular DP estimator for $\theta_0$, identifying it per Definition \ref{identification}. }

\textit{Another aspect that is worthwhile to illustrate in the context of this example is  a consequence of a ``non-strategic'' selection by a data curator within $\rbT_{\mathcal{E}}$. Generally, we should expect this to forfeit identification. For example, if the curator picks uniformly at random, the output $\theta(\rbT_{\mathcal{E}})$ has density
\[
\ell_{\theta(\rbT_{\mathcal{E}})}(t) = 
\begin{cases} 
0.5 / \theta_0 & \text{if } t < \theta_0, \\ 
0.5 / (M - \theta_0) & \text{if } t > \theta_0.
\end{cases}
\]
The non-degenerate of this distribution means $\theta_0$ is not identified per Definition \ref{nonidentification}. $\blacksquare$} 

Note that selecting a point uniformly at random from the random set $\rbT_{\mathcal{E}}$ in the limit or from $\rT_{N,\mathcal{E}}$ in the sample in the example above adheres to the `transparency" principle
of differential privacy (considered e.g. 
in \cite{gong:20}). Specifically, an external researcher can fully trace the data curator’s process from raw data to the final output. However, such a regular $  (\epsilon_N, 0)  $-DP (or more generally $(\epsilon_N, \delta_N)$-DP) output is neither a consistent estimator of the target parameter $  \theta_0  $ nor can it be transformed into one without additional information. Achieving consistent estimation in this setting requires the data curator to have complete knowledge of the distribution and possibly of the geometry of the  random set $\rT_{N,\mathcal{E}}$ and also analyze those of its weak limit $\rbT_{\mathcal{E}}$. 

The suggested approach of active collaboration with the data curator through the choice of a suitable strongly convex decision mapping is not the only way to restore point identification in the limit of DP outputs. Another mechanism that we believe will work in this example and select $\theta_0$ from the distribution of the random set $\rbT_{\mathcal{E}}$ is taking the quantile $Q_{r}(\rbT_{\mathcal{E}})$ for $r>0.5$ (for the definition of a random set quantile, see \cite{molchanov2005}).\footnote{\cite{khan2024sharp} noted that the quantile of a random set notion is useful for extracting a common boundary in realizations of the random set in the context of maximum score and estimator and some other semiparametric estimators. This idea is exactly relevant here as $\theta_0$ is a common boundary of the two realizations of the random set} The quantile approach does not fit our approach with the strongly convex function as its  loss function is convex but not strongly convex (and, thus, in general may not necessarily give a unique element in minimization). However,  we believe that with the right choice of the quantile index $r_N$ one could potentially accomplish $Q_{r_N}(\rT_{N,\mathcal{E}}) \stackrel{p}{\to} \theta_0$ quite generally. We leave this for future research.


A question remains whether strongly convex decision mappings with the properties discussed in this section and which allow us to identify $\theta_0$ in the limit can be constructed for general $\rT_{N,\mathcal{E}}$ with $\Theta$ having higher dimension than 1. We outline a general approaches for the case when the parameter of interest can be represented as a weighted integral over the random set $\rT_{\mathcal{E}}$

\paragraph*{$\theta$ can be represented as a weighted integral over the random set $\rT_{\mathcal{E}}$ using a fixed Lipschitz-continuous weight function.}

Let  $\rbT_{\mathcal{E}}^{\theta}$ denote the limiting random set of regular DP estimators corresponding to the 
data generating process with parameter $
\theta \in \Theta$.  Here we consider the case when \textit{there exists a fixed
continuous 
mapping $M(\cdot)$ defined by Lipschitz-continuous function $m(\cdot)$
such that for each realization of the random
set $\rbT_{\mathcal{E}}^{\theta},$} 
\begin{equation}\label{condition:1}
\theta=M\left( \rbT_{\mathcal{E}}^{\theta}\right)=
\int_{\rT_{\mathcal{E}}^{\theta}}\zeta\,m(\|\zeta\|^2)\,d\zeta
\end{equation}
with the normalization condition \(\int_{\rbT_{\mathcal{E}}^{\theta}} m(\|\zeta\|^2) \, d\zeta = 1\). This represents \(\theta\) as a weighted centroid of \(\rbT_{\mathcal{E}}^{\theta}\), where \(m(\|\zeta\|^2)\) assigns weights based on the squared norm of points, allowing \(\theta\) to lie anywhere within \(\rbT_{\mathcal{E}}^{\theta}\), including its boundary. 

This generalizes simpler selectors, such as the barycenter (where \(m(\|\zeta\|^2) = \frac{1}{\text{vol}(\rbT_{\mathcal{E}}^{\theta})}\)), used in econometric models of partial identification \citep{BeresteanuMolinari08,molinari2008partial}. The flexibility of \(m(\cdot)\) enables the approach to handle non-symmetric random sets. For instance, it is applicable to  our example in Section \ref{ex:samplemean2}, where \(\rbT_{\mathcal{E}}^{\theta} = [0, \theta]\) or \([\theta, M]\) with probability 1/2 and a tailored \(m(\cdot)\) can select \(\theta\) from both realizations,

To enhance computational tractability, we apply the divergence theorem to express \eqref{condition:1} as a surface integral. Define \(\mu(t)\) such that \(\mu'(t) = \frac{1}{2} m(t)\), with \(\mu(0) = 0\). For any fixed vector \(u \in \mathbb{R}^p\), we have $\langle \theta, u \rangle = \int_{\rbT_{\mathcal{E}}^{\theta}} \langle \zeta, u \rangle m(\|\zeta\|^2) \, d\zeta = \int_{\rbT_{\mathcal{E}}^{\theta}} \text{div} \left( \mu(\|\zeta\|^2) u \right) \, d\zeta$,
since $\text{div} \left( \mu(\|\zeta\|^2) u \right) = \sum_{i=1}^p \frac{\partial}{\partial \zeta_i} \left( \mu(\|\zeta\|^2) u_i \right) = 2 \mu'(\|\zeta\|^2) \sum_{i=1}^p \zeta_i u_i = m(\|\zeta\|^2) \langle \zeta, u \rangle$.

By the divergence theorem, this becomes $\langle \theta, u \rangle = \oint_{\partial \rbT_{\mathcal{E}}^{\theta}} \mu(\|\zeta\|^2) \langle u, dS \rangle$, where \(dS\) is the outward normal surface element on the boundary \(\partial S\rbT_{\mathcal{E}}^{\theta}\). Since this holds for all \(u\), we have:
\begin{equation}\label{divergence}
\theta = \oint_{\partial \rbT_{\mathcal{E}}^{\theta}} \mu(\|\zeta\|^2) \zeta \, dS.
\end{equation}
The normalization $\int_{\rbT_{\mathcal{E}}^{\theta}} m(\|\zeta\|^2) \, d\zeta = 1$ is enforced via a function \(\nu(t)\) satisfying \(\nu(0) = 0\) and $\text{div} \left( \nu(\|\zeta\|^2) \zeta \right) = \sum_{i=1}^p \frac{\partial}{\partial \zeta_i} \left( \nu(\|\zeta\|^2) \zeta_i \right) = m(\|\zeta\|^2)$. This implies the differential equation \(2t \nu'(t) + p \nu(t) = m(t)\), with solution $\nu(t) = \frac{1}{2} t^{-p/2} \int_0^t \tau^{p/2 - 1} m(\tau) \, d\tau$. Thus, normalization becomes $\oint_{\partial \rbT_{\mathcal{E}}^{\theta}} \nu(\|\zeta\|^2) \langle \zeta, dS \rangle = 1$.

The decision mapping is constructed as $f(z) = \left\| z - \oint_{\partial \rbT_{\mathcal{E}}^{z}} \mu(\|\zeta\|^2) \zeta \, dS \right\|^2$,
subject to normalization for $\mu(\cdot)$ so that 
corresponding density $m(\cdot)$ integrates
to 1. This mapping is strictly convex.  This approach leverages the boundary geometry of \( \rbT_{\mathcal{E}}^{\theta}\), aligning with econometric applications where the shape of identified sets encodes the parameter \citep{BeresteanuMolinari08}.

To approach this case computationally, the data curator requires (1) the ability to simulate realizations of \(\rbT_{\mathcal{E}}^{\theta}\) for \(\theta \in \Theta\), and (2) a representation of \(\rbT_{\mathcal{E}}^{\theta}\)’s boundary (e.g., via vertices of a polyhedral approximation). 

To approximate \(\mu(t)\), define a grid \(G = \{\theta^{(k)}\}_{k=1}^K \subset \Theta\). For each \(\theta^{(k)}\), simulate a realization \(T_{\theta^{(k)}}\) of \(\rbT_{\mathcal{E}}^{\theta_k}\). Represent \(\mu(t) \approx \hat{\mu}(t) = \sum_{r=0}^R \alpha_r h_r(t)\), where \(h_r(t)\) are orthogonal polynomials. 
This representation allows us to reduce the 
problem of finding function $\mu(\cdot)$
to the problem of finding $R+1$ coefficients
of its orthogonal representation. Then for each simulated realization $T_{\theta^{(k)}},$ we compute
$2(R+1)$ surface integrals
$
H_{rk}=\oint\limits_{\partial T_{\theta^{(k)}}}
h_r(\|\zeta\|^2)\,dS$, 
$Q_{rk}=\oint\limits_{\partial T_{\theta^{(k)}}}
q_r(\|\zeta\|^2)\langle \zeta,\,\,dS \rangle$, $r=0,\ldots,R$, 
where $q_r(t)=\frac12 t^{-p/2}\int^t_0 
\tau^{p/2-1}h^{\prime}_r
(\tau)\,d\tau.$
Coefficients $\widehat{\alpha}_r$ of the 
orthogonal representation are found
by solving the constrained quadratic 
optimization 
problem 
\begin{equation}\label{num:mapping}
\min\limits_{\alpha_0,\ldots,\alpha_R}
\sum\limits^K_{k=1}
(
\theta^{(k)}-\sum\limits^R_{r=0}\alpha_r\,H_r
)^2,\quad\quad
\mbox{subject to } \; \;\;
\frac{1}{K}\sum\limits^K_{k=1}\alpha_r
Q_{rk}=1.
\end{equation}
The constraint ensures that numerical
approximation for the density $\mu(\cdot)$
approximately integrates to 1 over the 
instances of $T_{\theta^{(k)}}.$ 

The resulting decision mapping is 
\[
\hat{f}(z) = \left\| z - \sum_{r=0}^R \hat{\alpha}_r \oint_{\partial \rT_{N,\mathcal{E}}} h_r(\|\zeta\|^2) \zeta \, dS \right\|^2.
\]
for a realization of the random set
$\rT_{N,\mathcal{E}}$.

\begin{lemma}\label{approximation:lemma}
The mapping \(\hat{M}(T) = \sum_{r=0}^R \hat{\alpha}_r \oint_{\partial T} h_r(\|\zeta\|^2) \zeta \, dS\), where \(\hat{\alpha}_r\) solve \eqref{num:mapping}, converges uniformly to \(M(T) = \int_T \zeta m(\|\zeta\|^2) \, d\zeta\) over compact sets in the Fell topology on \(\Theta\).
\end{lemma}

As mentioned above, this approach generalizes barycenter-based methods common in econometrics, where the Aumann expectation (barycenter) summarizes identified sets \citep{BeresteanuMolinari08,molinari2008partial}. Unlike the barycenter,  our method allows \(m(\cdot)\) to place \(\theta\) anywhere in \(\rbT_{\mathcal{E}}^\theta\), accommodating complex geometries as in Example in Section \ref{ex:samplemean2}. This flexibility enhances applicability of our method while ensuring differential privacy via post-processing \citep{dwork2014algorithmic}.

\section{Inference based on a single observation: Bayesian credible regions} 
\label{sec:bayes}

When the data curator does not collaborate to ensure point identification through a decision mapping, as outlined in Section \ref{sec:datacurator}, the researcher is left with a single differentially private (DP) output \(\theta(P_N, \nu_N) \in \rT_{N,\mathcal{E}}\). In such cases, the limiting random set \(\rbT_{\mathcal{E}}\) is often non-degenerate, implying that \(\theta_0\) is neither point-identified nor partially identified in the classical sense. This section proposes a Bayesian framework for finite-sample inference, constructing credible regions to quantify uncertainty about \(\theta_0\) based on this single DP output, while accounting for the randomization introduced by DP mechanisms.

The fact that  only a single output \(\theta(P_N, \nu_N) \in \rT_{N,\mathcal{E}}\) to a query is available (due to privacy budget constraints \citep[page 9]{dwork2014algorithmic}) renders  traditional frequentist confidence intervals infeasible, as they require repeated sampling. A Bayesian approach is particularly suitable here, leveraging all available information
(Including possible structure of the random set and the rule used by the data curator
to produce a differentially private
output)
to make the best possible inference
regarding the target parameter.  Priors can also reflect knowledge gained from publicly available datasets that do not require DP guarantees.  Bayesian analysis will yield  credible regions for $\theta_0$. Unlike classic confidence sets, these regions may not shrink to a point when the limiting random set $\rbT_{\mathcal{E}}$ is not a singleton. Nonetheless, credible regions can be constructed using procedures analogous to standard confidence intervals, making them practical for applied work even under DP constraints.

The Bayesian model utilizes the likelihood
function $\ell_{f,\rT_{N,{\mathcal E}}}(t; \theta)$ of the output of the decision
mapping constructed from its distribution
$\PP\left(
\tau_f\left(\rT_{N,\mathcal{E}}\,;
\,\theta\right) \leq t
\right)$ induced by the distribution 
of the random 
set $\rT_{N,\mathcal{E}}.$
We explicitly use index $\theta \in 
\Theta$ for the probability to indicate that $\rT_{N,\mathcal{E}}$ is the random
set of regular differentially private
estimators corresponding to the data
generating process indexed by $\theta.$ The  decision mapping now, of course, does not need to  be tailored  to the target parameter and satisfy requirements of Section .... , It  reflects the knowledge of a researcher of e choice made by the data curator (e.g. an elements could be chosen from $\rT_{N,\mathcal{E}}$ completely randomly). The Bayesian model then is determined
by the tuple $(\rT_{N,\mathcal{E}},
f(\cdot),\pi(\cdot))$ where $\pi(\cdot)$
is the prior of the researcher over the 
values of the parameter of the data
generating process.

The researcher combines the likelihood
function $\ell_{f,\rT_{N,{\mathcal E}}}(t;\theta)$ with the information from the 
prior to form the posterior distribution
for the target parameter given the realization $t=\tau_f\left(\rT_{N,\mathcal{E}}\right)$ of the 
decision mapping of the data curator:
\begin{equation}\label{sample:posterior}
\Pi_N(B\,;\,t)  \propto 
\int_{B}\ell_{f,\rT_{N,{\mathcal E}}}(t;\theta)\,
\pi(\theta)\,d\theta,
\end{equation}
where prior $\pi(\cdot)$ plays the role of weighting the
likelihood functions corresponding to data generating processes
indexed by a given $\theta$. By weak convergence of random sets
$\rT_{N,{\mathcal E}}$
to $\rbT_{{\mathcal E}},$
the posterior $\Pi_N(B\,;\,t)$ converges
pointwise to the limiting distribution
\begin{equation}\label{population:posterior}
\Pi(B\,;\,t)  \propto 
\int_{B}\ell_{f,\rbT_{{\mathcal E}}}(t;\theta)\,
\pi(\theta)\,d\theta,
\end{equation} 
as 
$N \rightarrow \infty.$ We then view
inference drawn from finite sample
prior $\Pi_N(B\,;\,t)$ as an approximation
for inference that would be drawn from the 
limiting distribution of the random set
$\rbT_{{\mathcal E}}$ and the
corresponding posterior $\Pi(B\,;\,t).$ 

Our proposed method to draw informative inference is an $(1-\alpha)$-level credible region, $\alpha\in (0,1)$, for $\theta_0$ which is the set $\cB_{\alpha}(t)$ with the property
$
\Pi\left(\cB_{\alpha}(t)\,;\,t \right) \geq 1-\alpha
$ 
and often the highest posterior density region to minimize volume. This region contains \(\theta_0\) with posterior probability \(1-\alpha\).\footnote{This concept of credible regions is standard in Bayesian statistics. See e.g. \cite{gelman2013bayesian}.}

{\bf EXAMPLE in Section \ref{ex:samplemean2} 
(continued)} 
Suppose that the data curator picks the 
point in the random set of regular
differentially private estimators uniformly at random and the researcher uses the uninformative prior.
Based on our previous discussion, the likelihood for the 
differentially private
output $t$ can then be expressed
as
$ 
\ell_{f,\rbT_{{\mathcal E}}} (t;\theta)=\frac{.5}{\theta} \mathbf{1}(t<\theta) + 
\frac{.5}{M-\theta} \cdot \mathbf{1}(t\geq \theta).
$ 
Then the posterior density can be expressed
as
$$
\ell_{f,\rbT_{{\mathcal E}}}(t;\theta)\,
\pi(\theta)=\left(
-\log\left(t(M-t)/M^2\right)
\right)^{-1}
\left(
(M-\theta)^{-1}\,{\bf 1}\{\theta<t\}+\theta^{-1}
\,{\bf 1}\{\theta \geq t\}\right).
$$
Thus, we can define credible region
$\cB_{\alpha}(t)=[z(t),
M-z(t)]$ such that 
$\Pi(\cB_{\alpha}(t);t)=
\int_{z(t)}^{
M-z(t)}\ell_{f,\rbT_{{\mathcal E}}}(t;\theta)\,
\pi(\theta)
\,d\theta=1-\alpha.
$ 
Solving this equation for $z(t)$ yields
$
z(t)=
 M^{1 - \alpha} [t(M - t)]^{\alpha/2}
$ and $\cB_{\alpha}(t)=[ M^{1 - \alpha} [t(M - t)]^{\alpha/2},\,
 M-M^{1 - \alpha} [t(M - t)]^{\alpha/2}]$ whenever
data curator produces
a differentially private
output $t.$ $\blacksquare$

The approach outlined in this section aligns with econometric traditions for handling weak or partial identification, where inference is possible despite identification failure. In instrumental variables (IV) models with weak instruments \citep{staiger1997instrumental,stock2005testing}, the first-stage $F$-statistic is low, leading to non-degenerate limit distributions and non-identifiability, similar to the non-degenerate \(\rbT_{\mathcal{E}}\) in DP. Bayesian methods in weak IV settings \citep{chernozhukov2008inference,kleibergen2005generalized} use priors to produce credible intervals that are robust to identification failure, paralleling our use of priors to form credible regions from a single DP output. In partial identification, Bayesian posteriors over identified sets yield credible regions \citep{moon2012bayesian,kline2016bayesian}. Our framework extends this to DP, where \(\rbT_{\mathcal{E}}\) acts as a "stochastic identified set," and credible regions quantify uncertainty when point identification fails due to privacy noise.

 \section[Average Treatment Effect]{Application to (Local) Average Treatment Effects Estimation}
\label{sec:appl}

Having established the frameworks for privacy-aware  identification, we now demonstrate its practical implications.

\subsection{Inverse Propensity Scores (IPS) estimation.} In this section, we apply concepts introduced earlier to the estimation of average treatment effects (ATE) using IPS estimators, illustrating how differential privacy affects identification in a common econometric setting.

The analysis of treatment effects focuses on identifying the impact of an intervention by comparing observed outcomes to counterfactual potential outcomes for the same unit. A key classical challenge is that both potential outcomes are never observed for any individual: we denote $Y_{1i}$ as the outcome if unit $i$ receives treatment ($D_i=1$) and $Y_{0i}$ if not ($D_i=0$). The observed outcome is $Y_i = D_i Y_{1i} + (1 - D_i) Y_{0i}$. Without loss of generality, assume the support of $Y_{1i}$ and $Y_{0i}$ is $[0,1]$. While individual treatment effects $Y_{1i} - Y_{0i}$ cannot be recovered, the average treatment effect (ATE), $\theta_0 = E[Y_{1i} - Y_{0i}]$, can be identified from the data under some conditions and is a common parameter of interest (assume $\theta_0 \in [0,M]$). We denote realizations of random variables by lowercase letters and the variables by uppercase.

One identification strategy, from \cite{rosenbaum:83} (see also \cite{hahn:98}, \cite{hirano:03}, \cite{abadie:06}), relies on unconfoundedness and a full support assumptions (we also maintain SUTVA without formulating it here explicitly): 

\begin{assumption} \label{ate assumption} 
Let the following hold:
\begin{description}
  \item[(i)] There exists an observed covariate $X_i$ such that $D_i \perp (Y_{0i}, Y_{1i}) \mid X_i$.
  \item[(ii)] $0 < P(D_i=1 \mid X_i) < 1$ for all $X_i$.
\end{description} 
\end{assumption}

Under Assumption \ref{ate assumption}, $\theta_0$ is identified as $\theta_0 = E_X[E[Y \mid D=1, X] - E[Y \mid D=0, X]]$.
Equivalently,
\begin{equation}\label{ateiden}
\theta_0 = E\left[\frac{Y(D - p(X))}{p(X)(1 - p(X))}\right],
\end{equation}
where $p(X) = P(D=1 \mid X)$ is the propensity score. This is a weighted moment condition, where the denominator shrinks if $p(X)$ nears 0 or 1. Identification fails if any region of $X$'s support $\mathcal{X}$ is trimmed (fixed trimming does not recover $\theta_0$).

We now characterize the limit of randomized regular
differentially private statistics $\theta(\PP_N,\nu_N)$ which are smooth (i.e., satisfy
Definition \ref{smoothness}). Since the parameter
of interest can be expressed as an expectation of
a weighted mean of outcome variable $Y,$ we can
use the intuition in our Example \ref{ex:samplemean2} which allowed us
to use a trimmed weighted sample mean to construct
an estimator which is both regular and differentially
private.

Under Assumption \ref{ate assumption}  the following IPS estimator from a random sample $\{(y_i,x_i,d_i)\}_{i=1}^N$ can be used for 
$\theta_0$ : 
$$\widehat{\theta}= \frac{1}{N} \sum_{i=1}^N \left(\frac{y_i d_i}{p(x_i)}- \frac{y_i(1-d_i)}{1-p(x_i)} \right),$$ 
where for simplicity of exposition we
use the true value of the propensity score $p(\cdot)$ to form the weights for $y_i$ rather than its feasible estimnate $\widehat{p}(x_i)$. 

Without {\it a priori} distributional knowledge, a
researcher would not know if the fact that the propensity score function can vary between $0$
and $1$ should be treated as the only constraint, or
there is some possibly small constant that bounds
the sample values of the propensity score
away from 1 and 0. In the context where the 
randomized statistic representing the target
parameter $\theta_0$ needs to be evaluated without 
such a prior knowledge and yet satisfy DP, the unbounded range of the elements of the sum in the weighted mean representing $\widehat{\theta}$ need to be trimmed to guarantee
the boundedness of the support of each element.

To do so, we can introduce a sequence
$h_N \rightarrow 0$ of the trimming parameters
and an ``effective sample size" $n_N=N\,\PP(h_N \leq p(X) \leq 1-h_N)$ such that the trimmed
estimator can be written as 
$\widetilde{\theta}=   n_N^{-1}\sum_{i,\,
h_N \leq p(x_i) \leq 1-h_N}  \left(y_i d_i \big/p(x_i)- y_i(1-d_i)\big/(1-p(x_i)) \right).$ 
This is an estimator frequently used by the 
practitioners in the absence of clear {\it a priori}
bound for the propensity score.

One significant advantage of the estimator $\widetilde{\theta}$ is that the support
of each element in the sum is contained in the
interval
$[-M/(n_N(1-h_N)),\,M/(n_N h_N)].$
This also means that we can construct a 
differentially private estimator by using
simple additive mechanisms discussed above.
Take $\theta(\PP_N,\nu_N)$ to be equal to
$
\widetilde{\theta}
+\widetilde{a}(\nu_N),
$
where $\widetilde{a}(\nu_N)$ is the Laplace random variable
with mean zero and 
parameter $1/(n_Nh_N(1-h_N)\varepsilon_N),$ if 
$
\widetilde{\theta}
+\widetilde{a}(\nu_N),
$ falls within the interval $[0,M].$ If the sum falls
outside of $[0,M],$ $\theta(\PP_N,\nu_N)$ is set to be the closest
end of the interval.
Invoking a classic Theorem 1 in \cite{dwork:2006},
we can verify that randomized statistic
$\theta(\PP_N,\nu_N)$ is $(\varepsilon_N,0)$-differentially private.

The effective sample size under truncation $n_N$ is generally smaller than $N$. Of course, to guarantee that the trimmed estimator $\widetilde{\theta}$
converges in probability to the target parameter
$\theta_0$, it is necessary that the trimming sequence vanishes in the limit, $h_N \rightarrow 0$. However, this condition alone is not sufficient
for $\widetilde{\theta}$ to have a degenerate
weak limit at $\theta_0.$  This guarantee cannot be 
provided unless there is {\it a priori knowledge} regarding the distribution of the propensity 
score $p(X).$ Without that knowledge and even knowing that $h_N \to 0$, the following
 regimes are possible, echoing Example \ref{ex:samplemean2} (for illustrational simplicity , we disregard an analogue of Regime 3 in Example \ref{ex:samplemean2}, consider only the Laplace mechanism and a constant $\varepsilon_N$):

{\bf Regime 1: $n_Nh_N(1-h_N) 
\rightarrow 0$.} The variance
of the noise in the Laplace
noise diverges to infinity $(\varepsilon n_N h_N(1-h_N))^{-2} \rightarrow \infty.$
Due to the projection of the parameter value on
the parameter space $[0,M]$, the 
randomized statistic $\theta(\PP_N,\nu_N)$
converges in distribution
to a Bernoulli random variable taking values $0/M$
with probabilities $1/2$.

{\bf Regime 2: $n_Nh_N(1-h_N) 
\rightarrow \infty$.} 
The variance
of the Laplace noise converges
to $0$ as $N \rightarrow \infty$.
Since  $h_N \rightarrow 0$ ensures that 
$\widetilde{\theta} \stackrel{p}{\to}\theta_0$, then the randomized statistic $\theta(\PP_N,\nu_N)$
converges in probability to the 
target parameter.

Next, note that the estimator projected on 
$[0,M]$
can be approximated by $\widetilde{\theta} -\theta_0+
\min\{M,\max\{0,\theta_0+\widetilde{a}(\nu_N)\}\}$ up to a term that converges in probability to 0. We can represent
$\widetilde{a}(\nu_N)=F^{-1}_{\Lambda}(\nu_N)/
(n_Nh_N(1-h_N)\varepsilon),$ where
$F_{\Lambda}(\cdot)$ is the distribution function
of the standard Laplace distribution. $\widetilde{\theta}-\theta_0$ constructed from $\widetilde{\theta}$ above is a Lipschitz functional
of the empirical distribution (as a weighted
sample mean) and it converges weakly to zero in both Regimes 1 and 2. Function $\min\{M,\max\{0,\theta_0+\widetilde{a}(\cdot)\}\}$
converges in both regimes with upper and lower
envelopes bounded by 0 and $M$. This means that 
randomized statistic $\theta(\PP_N,\nu_N)$
is regular as in Definition \ref{regularity} and is smooth as in Definition \ref{smoothness} for both
Regime 1 and Regime 2. In the  Definition  \ref{smoothness} we can take $\psi_N(\PP_N)=\widetilde{\theta}-\theta_0$
and $a(\nu_N)=\min\{M,\max\{0,\theta_0+\widetilde{a}(\cdot)\}\}$. 
As a result, we can apply Theorem \ref{representation:smooth} where the set of
weak limits of $\psi(\PP_N)$ is a singleton $\{0\}$ and the set of weak limits of 
$a(\nu_N)$ must be the superset of the convex
envelope of a degenerate random variable at $\theta_0$ and $B(.5),$ a Bernoulli random variable taking values $0/M$ with probability $.5.$

Thus, the limiting random set is 
$\rbT_{\mathcal E}=\Psi \oplus {\mathcal A},$
where $\Psi=\{0\}$
and ${\mathcal A}$ is a random set with a non-degenerate distribution. In other
words, the limiting random set $\rbT_{\mathcal E}$ has non-generate distribution and, thus,  neither point nor partially identifies the ATE.

\subsection{Regression discontinuity design (RDD)} 
\label{sec:RDDdetail} 

Regression discontinuity design (RDD) is one of the most widely used quasi-experimental methods for causal inference, and it is often viewed as a particularly credible identification strategy (see \cite{Hahnetal:01}, \cite{ImbensLemieux}, \cite{LeeLemieux},  \cite{cattaneo_idrobo_titiunik_2020}). In this section we illustrate how DP affects identification in both the sharp and fuzzy RDD settings. As we will see, the main identifcation challenge arises not so much from the unknown distributional properties driving tuning parameters but rather the traditional nonparametric estimators in this literature  being sensitive to changes in individual observations. 

A detailed treatment, including formal derivations and simulation evidence, is provided in Appendix B.

Let $Y_i$ denote the outcome, $D_i \in \{0,1\}$ the treatment indicator, and $X_i$ the running variable. In the sharp design treatment assignment is deterministic $D_i = 1\{X_i \geq c\}$ for some known cutoff $c$. The average causal effect is identified as the jump in conditional expectation of $Y$ at the cutoff:
\[
\theta_{0,S} = \lim_{x \downarrow c} E[Y|X=x] - \lim_{x \uparrow c} E[Y|X=x].
\] 
In the fuzzy design, under the assumption that the treatment probability exhibits a discontinuity at $c$, we identify 
\[
\theta_{0,F} = \frac{ \lim_{x \downarrow c} E[Y|X=x] - \lim_{x \uparrow c} E[Y|X=x]}{ \lim_{x \downarrow c} P(D=1|X=x) - \lim_{x \uparrow c} P(D=1|X=x)}.
\]
Under standard assumptions in this literature, standard nonparametric methods such as kernel regression at the boundary or local linear regression estimate these quantities consistently under a suitable choice of the bandwidth sequence $h_N$ with $h_N \rightarrow 0$ as $N \rightarrow \infty$. There are some well-known and widely used approaches for selecting a bandwidth, such as \cite{ImbensK}, \cite{Calonicoetal2014ECMA}, among others. 

We focus on the class of smooth regular DP estimators. Smoothness (or approximate additivity), as mentioned above, to the best of our knowledge describes all common practical DP methods. Smooth estimators include a subclass of additive estimators 
$\widehat{\theta}+a(\nu_N)$ (e.g., obtained by means of Laplace or Gaussian mechanisms described earlier) projected onto the parameter space. The weak limit of the random set $\rT_{N,\mathcal{E}}$ of such estimators by Theorem \ref{representation:smooth}  has Minkowksi representation. Our finding for the subclass of additive estimators  will apply to \textit{the  class of all smooth estimators in this econometric context}. We focus on  fully (and not just approximate) additive estimators for illustrational simplicity. 

Our results in Appendix B on possible maximum changes in the original non-private estimator $\widehat{\theta}$ (whether local linear or nonparametric mean at the boundary) imply that a randomized statistic $\widehat{\theta}+a(\nu_N)$ 
will have  poor limiting properties as the additive noise 
$a(\nu_N)$ needs to be calibrated by the worst-case
performance of the estimator $\widehat{\theta}$.  Essentially, since $\widehat{\theta}$ can have large variations on the real line with the change in one observation, in order to provide the DP
guarantee,
the variance of $a(\nu_N)$ does not approach zero as 
$N \rightarrow \infty$ as long as $\epsilon_N$ in the $(\epsilon_N,\delta_N)$-differential privacy guarantee remains bounded from above. (This is discussed more formally in Appendix B.) With bounded $\varepsilon_N$, randomized statistics $\widehat{\theta}+a(\nu_N)$ with a proper choice of bandwidth sequence (that guarantee consistency of non-private $\widehat{\theta}$) will have non-degenerate weak limits driven by the asymptotic distribution of $a(\nu_N)$ leading to the weak limit $\rbT_{\mathcal E}=\{\theta_0\} \oplus {\mathcal A}$ of $\rT_{N,\mathcal{E}}$ being non-degenerate due to ${\mathcal A}$ not being $\{0\}$. These finding are summarized in Theorem \ref{th:RDDgeneral}.  

\begin{theorem}
\label{th:RDDgeneral}
Under standard RDD conditions for either sharp or fuzzy design, a class of smooth regular DP estimators that build on either nonparametric regression at the boundary or local linear estimators   does not point identify the casual effect in the limit of statistical experiments if the sequence of privacy budgets $\epsilon_N$ remains bounded. As a result, the limiting random set $\rbT_{{\mathcal E}}$ has a non-degenerate distribution.
\end{theorem} 

Thus, our general conclusion for a class of smooth DP nonparametric RDD estimators is the failure to identify the treatment effect in the limit of statistical experiments. If we add other covariates to our estimation,  the qualitative conclusions on the lack of identifiability of the parameter of interest attained in Theorem \ref{th:RDDgeneral} will remain the same. 
Thus, DP requirements here disrupt the concentration of nonparametric estimators, producing limits that are random sets

There are important lessons that emerge in the RDD setting. First, because RDD estimators rely on shrinking neighborhoods around the cutoff, privacy noise does not diminish asymptotically as it might for global estimators. We have, therefore, a reason to believe that similar issues will arise in other applied econometrics approaches that rely on ``thin sets'' for identification and estimation in the absence of privacy noise. Second, the data curator involvement would be essential as with curator-based decision mappings (see Section \ref{sec:datacurator}), identification may be restored.

RDD example is complementary to our earlier analysis of inverse propensity score estimatorsand it demonstrates that DP requirements fundamentally alter identification in leading econometric designs. In both cases, the main conclusion is the same: unless the data curator actively exploits the structure of the random set of DP estimators, parameters that are point-identified in the non-private world may fail to be identified under DP. Further theoretical derivations and discussions of specification tests for RDD are presented in Appendix B.

\vskip 0.05in

\textit{Monte Carlo illustration} We illustrate our findings of a generally poor performance of the DP RDD estimators obtained by means of additive mechanisms (or, more generally. smooth mechanisms). We consider the sharp design and illustrate paths of the differentially private local linear estimator with a triangular kernel for increasing sample sizes with different degrees of the privacy protection. These paths are constructed for increasing samples from the size of 300 till the size of 4000. For illustrational simplicity, we give paths for 20 independent realizations of datasets without projecting them on the parameter space (which is the practice we refer to in our theoretical exposition in the main paper). The Monte Carlo scenario  is inspired by a design in \cite{ImbensK}.

Take the forcing variable $X$ to have a uniform distribution on $[-1,1]$. The regression function is a fifth-order polynomial, with separate coefficients for
$X_i <0$ and $X_i >0$: 
$$
m(x)= \left\{
\begin{array}{l}
0.35+1.27x + 7.18x^2+20.21x^3+21.54x^4+7.33x^5, \quad \text{if } x < 0,	\\ 
0.65+0.84x-3x^2+7.99x^3-9.01x^4+3.56x^5, \quad \text{if } x \geq 0, 
\end{array}	
\right.
$$
and the error $u$ having a symmetric uniform distribution on $[-0.12952 \cdot \sqrt{3}, 0.12952 \cdot \sqrt{3}]$. The bandwidth in the local linear estimation is chosen using the approach in \cite{ImbensK}.  DP  estimators are obtained by using the Laplace mechanism, which draws a mechanism nose from the Laplace distribution with mean zero and the variance calibrated to ensure the desired level of privacy. 

\begin{sidewaysfigure}[ht]
    \includegraphics[width=0.95\textwidth]{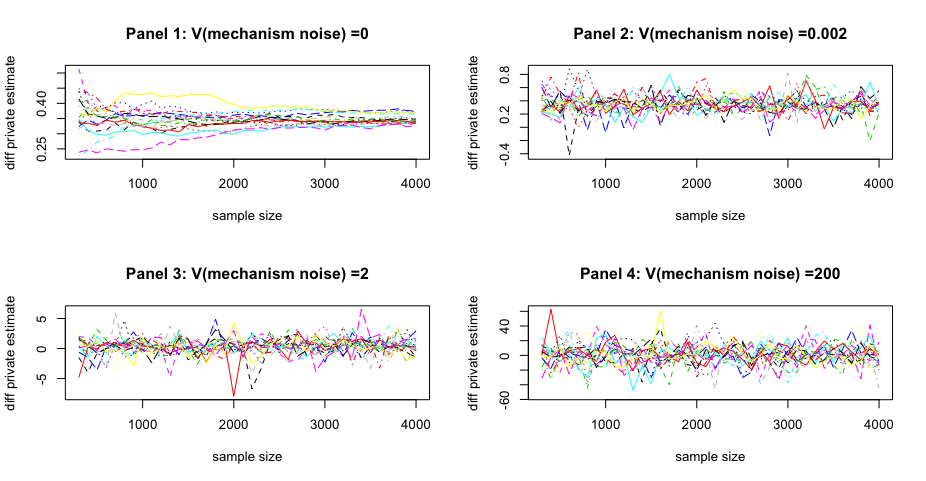}
	\caption{RDD illustration. Twenty independent paths of differentially private estimators local linear estimators for increasing sample sizes for various degrees of differential privacy protection.}
	 \label{fig:FigureMC1} 	 
\end{sidewaysfigure}

Panel 1 in Figure \ref{fig:FigureMC1} shows the paths of the estimator in the absence of the mechanism noise. Panel 2 in Figure \ref{fig:FigureMC1} depicts the paths of the estimator when the mechanism noise variance equals $0.002$ for any $N$ -- this corresponds to $\varepsilon_N$ being 10 times of $4 \cdot 0.12952 \cdot \sqrt{3}$ and $\delta_N=0$.\footnote{The data curator can take supports of the treatment outcome to be $[0.35-0.12952 \cdot \sqrt{3}, 0.35+0.12952 \cdot \sqrt{3}]$ to the left of the cut-off and $[0.65-0.12952 \cdot \sqrt{3}, 0.65+0.12952 \cdot \sqrt{3}]$ to the right of the cut-off. This choice of supports is most favorable to the DP approach.}  Panel 3 in Figure \ref{fig:FigureMC1} shows the paths when the mechanism noise variance equals  $2$ for any $N$ (corresponds to $\varepsilon_N=4 \cdot 0.12952 \cdot \sqrt{3}$ and $\delta_N=0$). Finally, Panel 4 in Figure \ref{fig:FigureMC1} illustrates the paths when the mechanism noise variance equals   $200$ for any $N$ (this corresponds to $\varepsilon_N =0.1\cdot 4 \cdot 0.12952 \cdot \sqrt{3}$ and $\delta_N=0$). Please note the different range of the values on the vertical axis in these panels.

In Table \ref{table:MC1} we focus on the rejection of the null $H_0: \theta_{0,S}=0$ against $H_1: \theta_{0,S} \neq 0$ when a researcher uses differentially private estimates and their standard errors.

\begin{table}
\par
\begin{center}
\begin{tabular}{lllllllll}
\hline
Setting  \hspace*{0.2in} & \multicolumn{2}{c}{Var(mech noise =0)} & 
\multicolumn{2}{c}{Var(mech noise) =0.002} & \multicolumn{2}{c}{Var(mech noise) =2} & 
\multicolumn{2}{c}{Var(mech noise) =200} \\ 
&   5\% & 1\% & 5\% & 1\% & 5\% & 1\% & 5\% & 1\% \\ \hline
$N=500$ & 1 & 1 & 0.6846 &  0.3706 & 0.0666 & 0.0286 & 0.056  & 0.023 \\
$N=2000$ & 1 & 1 & 0.7252 &  0.3880 & 0.0664 & 0.0290 & 0.0666  & 0.0272 \\
$N=5000$ & 1 & 1 & 0.7260 &  0.3844 & 0.0694 & 0.0284 & 0.0594  & 0.0250 \\
 \hline
\end{tabular}%
\end{center}
\caption{{\protect\small Rejection
rates in 5000 simulations of the false null hypothesis $H_0:\theta_{0,S}=0$ in Scenario 1. $N$ denotes the number of observations.}}
\label{table:MC1}
\end{table}

Full details including proofs  are in Appendix B.

These illustrations reinforce the paper's core message: DP preserves privacy but necessitates random set theory for identification, with point recovery possible via curator-researcher collaboration.

\section{Conclusion}
\label{sec:conclusion}

Differential privacy  provides a rigorous framework for protecting data, preventing adversaries from inferring sensitive information or identifying individuals. It achieves this through randomized estimators, where independent mechanism noise ensures formal privacy guarantees.

This paper examined econometric identification under DP constraints. We showed that even in relatively simple settings, identification requires tools from random set theory. By studying sequences of DP estimators applied to growing datasets, we defined identification in terms of the set of weak limits of these estimators. Under mild regularity conditions, this limiting set is a convex, compact random set that must be described probabilistically, for example through its containment functional. Our findings suggest that the loss of point identification under DP may be intrinsic to many econometric models that depend on local or nuisance parameters, or that involve significant informational asymmetry between the data curator and the researcher.

We also discussed how point identification might be restored through more active involvement of the data curator who is aware of the identification challenges created by DP-induced noise and informational asymmetry. The major takeaway from our results is that, in light of the many ongoing practical efforts to implement privacy-preserving mechanisms for data analysis, it is vital to engage more deeply with the econometric community. Their expertise will be essential in developing credible methods that balance privacy protection with meaningful econometric insight.

\bibliographystyle{ecta}
\bibliography{nsf_dcm.bib}

\section[Appendix]{Appendix: Proofs for Section  \ref{sec:privacyaware}.} 

\textbf{Proof of Lemma \ref{lemma:TNconvexcompact}. } Let $\omega_{\nu}$ be the element of
the $\sigma$-algebra $\mcF_{\nu}$ associated with the random 
element $\nu_N$ and $\omega_{S}$ be the element of the 
$\sigma$-algebra $\mcF_{\mcZ}$ of the subsets of $\mcZ^n.$

The set \(\rT_{N, \mathcal{E}}\) depends on \((\omega_S, \omega_{\nu}) \in \Omega\) through the random empirical distribution \(\PP_N(\omega_S)\) and random element \(\nu_N(\omega_{\nu})\). A random closed set in a Polish space \(\mathbf{L}_1(\PP)\) is a measurable map from \((\Omega, \mathcal{F}, \mathbb{P})\) to \(\mathcal{F}(\mathbf{L}_1(\PP))\), the space of closed subsets equipped with the Effros \(\sigma\)-algebra. Here $\mathcal{F} = \mcF_{\mcZ}\times \mcF_{\nu}$ and $\mathbb{P}$  is the product probability measure.

For each \((\omega_S, \omega_{\nu}) \in \Omega\),  the set of values $\theta(\PP_N(\omega_S),
\nu_N(\omega_{\nu}))$ in $\rT_{N,\mathcal{E}}(\omega_S, \omega_{\nu})$ is closed by construction and bounded due to the boundedness of $\theta$. Therefore,  it is compact. Thus, if we now establish the measureability of $\rT_{N,\mathcal{E}}$, it will mean that $\rT_{N,\mathcal{E}}$ is a compact random set. We verify Molchanov’s condition for measurability by showing that the graph of \(\rT_{N, \mathcal{E}}\),
\[
\operatorname{Gr}(\rT_{N, \mathcal{E}}) := \left\{ ((\omega_S,\omega_{\nu}), \theta) \in \Omega \times \mathbb{R}^p : \theta \in \rT_{N, \mathcal{E}}((\omega_S,\omega_{\nu})) \right\} 
\]
is measurable. This holds because: (i) each \(\theta(\cdot, \cdot)\) in \(\rT^*_{N, \mathcal{E}}\) is measurable by Definition ~\ref{regularity}; (ii)  \({\bf L}_1(\PP)\) is separable and complete, and closure in this space preserves measurability. Thus,  $\rT_{N, \mathcal{E}}$ is a compact random set.

If  $\mathcal{E}$ is a join-semilattice in the coordinate-wise partial order for $(\varepsilon_N,\delta_N)$ -- that is, the join of any two sequences from $\mathcal{E}$ is also in $\mathcal{E}$, then we can generally consider $\rT_{N, \mathcal{E}}$ as a convex random set in the sense of Definition 4.32 in \cite{molchanov2005}. Consider $\theta(\PP_N,\nu_N)$
and $\theta'(\PP_N,\nu_N)$ that are realizations of two regular $(\epsilon_N,\delta_N)$-differential private and $(\epsilon_N',\delta_N')$-differential private estimators, respectively,  which belong to $\rT_{N, \mathcal{E}}$ (hence,  sequences of $\{(\epsilon_N, \delta_N)\}$ and $\{(\epsilon_N',\delta_N')\}$ are in $\mathcal{E}$), then by continuous mapping theorem their convex combination satisfies (\ref{rate}) and, thus, condition (ii) in Definition \ref{regularity}. Also, any convex combination $\tau \theta(\PP_N,\nu_N) + (1-\tau) \theta'(\PP_N,\nu_N)$ is a realization of the estimator $\tau \theta(\cdot,\cdot) + (1-\tau) \theta'(\cdot,\cdot) $. This estimator is differentially private by DP composition theorem for the sequence of $\{\left(\max\{\epsilon_N, \epsilon_N'\}, \max\{\delta_N, \delta_N'\}\right)\}$ which belongs to $\mathcal{E}$ by our assumption of $\mathcal{E}$ being a join-semilattice. Also note that the estimator $\tau \theta(\cdot,\cdot) + (1-\tau) \theta'(\cdot,\cdot) $ has a weak limit from the continuous mapping theorem as it is straightforward to show that $(\theta,\theta')^{T}(\cdot,\cdot)$ has a joint weak limit. This would satisfy condition (iii) in Definition \ref{regularity}. Finally, as we mentioned, some additional information about the DP mechanism may be available  but if this additional information about the DP mechanism permits both $\theta(\cdot,\cdot)$ and $\theta'(\cdot,\cdot)$ then there is no reason to think that the additional information would eliminate some of their convex combinations. 
Thus, we can take random set $\rT_{N,\mathcal{E}}$ to be convex. $\blacksquare$

\vskip 0.1in 

\textbf{Proof of Lemma \ref{lemma:weaklimit}:} Assume, contrary to the statement of the Lemma that $\theta(\PP_N,\nu_N) 
\stackrel{p}{\longrightarrow} \tau.$ This implies 
$\Delta_N=\theta(\PP_N,\nu_N)-\tau 
\stackrel{p}{\longrightarrow}0.$ Then $\theta(\PP_N,\nu_N)=\tau+
\Delta_N,$ and because $\tau$ is not constant, then conditional on  $\PP_N$ and $\PP_{N+1}$, 
estimator $\theta(\PP_N,\nu_N)$ and $\theta(\PP_{N+1},\nu_{N+1})$
cannot be independent. This, in its turn, will contradict the independence of elements $\nu_N$
and $\nu_{N+1}$, which is a fundamental requirement for DP. $\blacksquare$

\vskip 0.1in
 \textbf{Proof of Theorem \ref{convergence}. }
By design $\theta(\PP_N,\nu_N)$ is a measurable selection of the random set $\rT_{N,{\mathcal E}}.$ Set $cl\{\theta(\PP_N,\nu_N),\,\nu_N \in {\mathcal E}\}$ is the Castaing representation of the set $\rT_{N,{\mathcal E}}$ (see definition 2.14 in \cite{molchanov2005}).

For any finite collection of functions $\{\theta^{(1)}(\cdot,\cdot),\ldots,\theta^{(K)}(\cdot,\cdot)\}$ satisfying Assumption \ref{operators}  and Definition \ref{regularity} the vector of random variables
$\left(\theta^{(1)}(\PP_N,\nu_N),\ldots,\theta^{(K)}(\PP_N,\nu_N)\right) 
$
converges weakly jointly to a random vector $(\tau^{(1)},\ldots,\tau^{(K)})$ since all elements $\theta^{(k)}(\cdot,\cdot)$ have a distribution with a singular measure over $(\PP_N,\nu_N).$
As a result, the finite-dimensional distribution of the distance $\rho(x,\,\rT_{N,{\mathcal E}})=\|x-\theta(\PP_N,\nu_N)\|$ for $x \in \Theta$ and a given $\theta(\PP_N,\nu_N)$ converges to the finite-dimensional distribution
of $\rho(x,\,\rbT_{\mathcal E})=\|x - \tau\|$ where $\tau$ is a weak limit of $\theta(\PP_n,\nu_N).$ 
Set $\rbT_{\mathcal E}$ is convex, bounded and closed 
provided that the weak limit of the convex combination of measurable selections in 
$\rT_{N,{\mathcal E}}$ is equal to the convex combination of their limits.
Convergence of distributions of distance functions then implies that the support function of $\rT_{N,{\mathcal E}}$ converges weakly to the support function of set $\rbT_{\mathcal E}.$ By Proposition 6.13 in \cite{molchanov2005}, this means that random set $\rT_{N,{\mathcal E}}$ converges weakly random set $\rbT_{\mathcal E}.$ $\blacksquare$

\vskip 0.2in 

\textbf{Proof of Lemma \ref{lemma:DCmeasselec}.}
Since $\rT_{N,{\mathcal E}}(\omega)$ is convex and compact, $\arg\min_{z \in \rT_{N,{\mathcal E}}(\omega)} f(z) \subset \rT_{N,{\mathcal E}}(\omega)$. $\alpha$-strong convexity of $f(\cdot)$ ensures this argmin is a singleton by compactness of $\rT_{N,{\mathcal E}}$ it is an element of $\rT_{N,{\mathcal E}}$. 
Then by measurability of infinum Theorem 2.27 (ii) in \cite{molchanov2005}, $\tau_f(\rT_{N,{\mathcal E}})$ is a random variable, as a singleton random closed set.
 $\blacksquare$

\vskip 0.05in 

\textbf{Proof of Theorem \ref{th:DCconsistency}.}
The result follows directly from Theorem \ref{th:DCweakconv} and the equivalence of weak convergence and convergence in probability for degenerate limits. $\blacksquare$

\vskip 0.1in 

\textbf{Proof of Lemma \ref{approximation:lemma}.}  
By Theorem III.3 in 
\cite{vladimirov:76} the linear partial
differential equation 
$
\sum^p_{i=1}\frac{\partial (z_i\nu(z))}{\partial z_i}=m(z)
$
has a solution $\nu(\cdot)$ which is unique
for each Lipschitz-continuous function 
$m(\cdot)$ given the initial condition and similarly the ordinary differential 
equation $\mu'(z)=\frac12 m(z)$ has a unique solution
given the initial condition. Moreover, these solutions
are also Lipschitz-continuous. From the divergence
theorem (\ref{divergence}), mapping
$M(\cdot)$ in (\ref{condition:1})
can be represented via a surface integral.
By uniqueness of a representation of $m(\cdot)$
via $\mu(\cdot),$ this surface integral 
representation is also unique.

Finally, by Lipschitz-continuity of $\mu(\cdot)$ by Weirstrass' theorem it
can be approximated uniformly by a high-degree polynomial
with approximation error decreasing as 
the degree of the polynomial grows.
$\blacksquare$

\section{Appendix B: RDD setting}

Theorem \ref{th:RDDgeneral} follows from a series of lemmas that establish large deviations of nonparametric regression at the boundary estimators with the change in one observation and its implication for smooth regular DP mechanisms.

We start by reviewing the definition of nonparametric estimators in the RDD setting. First, a \textit{nonparametric regression at the boundary} estimator  in both sharp and fuzzy RDD.is defined as 
\begin{equation}\label{eq:estimator_nprb}\widehat{\theta} = \sum_{X_i \geq c} Y_i w_{i,r} - \sum_{X_i < c} Y_i w_{i,l}, \quad \text{where} 
\end{equation}	
$$
w_{i,r} = \frac{K\left((X_i-c)/{h_N}\right)}{\sum_{X_i \geq c} K\left((X_i-c)/h_N \right)}, \quad 
w_{i.l} =  \frac{K\left({(X_i-c)}/{h_N}\right)}{\sum_{X_i < c} K\left({(X_i-c)}/{h_N}\right)}.
$$
This estimator is quite intuitive, but it has well-known drawbacks particularly with respect to the bias term being linear in bandwidth, as discussed in \cite{Hahnetal:01}, \cite{Porter03} and \cite{ImbensLemieux}, among others. A \textit{loocal linear (more generally polynomial) regression estimator} is defined as follows. In the sharp design, this method conducts two optimization problems by fitting linear regression functions 
\begin{align}(\widehat{\alpha}_L, \widehat{\beta}_L) & = \arg \min_{\alpha_L, \beta_L} \sum_{i: X_i<c} K\left({(X_i-c)}/{h_N} \right)(Y_i -\alpha_L -\beta_L (X_i-c))^2, \label{def:LL_left} \\
(\widehat{\alpha}_R, \widehat{\beta}_R) & = \arg \min_{\alpha_R, \beta_R} \sum_{i: c \leq  X_i} K\left({(X_i-c)}/{h_N} \right) (Y_i -\alpha_R -\beta_R (X_i-c))^2, \label{def:LL_right}	
\end{align}	
and then estimating $\theta_{0,S}$ as $\widehat{\theta}= \widehat{\alpha}_R - \widehat{\alpha}_L$. 	
The asymptotic properties of this estimator can be found e.g., in \cite{Hahnetal:01}, among others, and they are based on the theory in \cite{Fan:92} and \cite{FanGijbels:96}. In  the fuzzy design it is defined analogously but with the use of IV techniques that rely on employment location indicators $\mathbf{1}(X_i\geq c)$ (potentially involved in some functional forms with $X_i$). 
 
We next analyze the identifiability of the treatment effect from a subclass of all regular differentially private estimators that build on the nonparametric regression at the boundary or local linear approaches given by Theorem \ref{th:RDDgeneral}. 

\subsection{Global sensitivity of the nonparametric regression at the boundary estimator}

Lemma \ref{lemma:weightedmean} and it extension take steps toward obtaining results for a maximal absolute change in a weighted average when one observation changes -- this maximal absolute change is commonly known in the DP literature as \textit{global sensitivity}. Lemma \ref{lemma:weightedmean} considers two weighted averages with the same number of components and with weights formed as in the nonparametric regression at the boundary estimator. Its extension covers cases when one averages has one more component. 

\begin{lemma}
	\label{lemma:weightedmean}
Consider two weighted averages	
\begin{align*}
q_1&=\sum_{i=1}^T w_i a_i + w_{T+1} a_{T+1}, \quad q_2=\sum_{i=1}^T \tilde{w}_i a_i + \tilde{w}_T \tilde{a}_T, \quad \text{ with } \\
 w_i&= \frac{b_i}{\sum_{i=1}^{T+1} b_i}, \quad  \tilde{w}_i =  \frac{b_i}{\sum_{i=1}^{T} b_i +\tilde{b}_{T+1}}, \quad i=1, \ldots, T,	\quad \tilde{w}_{T+1} =  \frac{\tilde{b}_{T+1}}{\sum_{i=1}^{T} b_i +\tilde{b}_{T+1}}. 
\end{align*}	
Suppose for some $c_2>c_1 \geq 0$ and $d_2>d_1$, 
\begin{align}
 c_1 & \leq (\text{ or} <) \,b_i , \; \tilde{b}_i \leq c_2, \quad i=1, \ldots, T+1, \label{cond1}\\	
d_1 & \leq a_i \leq d_2, \quad i=1, \ldots, T+1.	\label{cond2}
\end{align}	
\begin{enumerate}
\item[(a)] If $c_1=0$ and $|d_1|, |d_2| < \infty$, then 
$$\max_{a_1, \ldots, a_T, a_{T+1}, \tilde{a}_{T+1}, b_1, \ldots, b_T, b_{T+1}, \tilde{b}_{T+1} \text{ s.t. } (\ref{cond1}), (\ref{cond2})}\left|q_1-q_2 \right| = d_2-d_1.$$
\item[(b)] If $c_1>0$ and $|d_1|, |d_2| < \infty$, then 
$$\max_{a_1, \ldots, a_T, a_{T+1}, \tilde{a}_{T+1}, b_1, \ldots, b_T, b_{T+1}, \tilde{b}_{T+1} \text{ s.t. } (\ref{cond1}), (\ref{cond2})}\left|q_1-q_2 \right| = \frac{c_2 (d_2-d_1)}{T \cdot c_1+c_2}.$$
\item[(c)] If $d_1=-\infty$ or $d_2=+\infty$, then  
$$\max_{a_1, \ldots, a_T, a_{T+1}, \tilde{a}_{T+1}, b_1, \ldots, b_T, b_{T+1}, \tilde{b}_{T+1} \text{ s.t. } (\ref{cond1}), (\ref{cond2})}\left|q_1-q_2 \right| =+ \infty.$$
\end{enumerate}
In cases (a)-(c), $\max\left|q_1-q_2 \right| $ can be attained by a positive change as well as by a negative change  -- i.e., there are values of $a_t$'s, $b_t$'s and $\tilde{a}_{T+1}$, $\tilde{b}_{T+1}$ such that 
$q_1-q_2=\max\left|q_1-q_2 \right| $, and there are values such as $q_1-q_2=-\max\left|q_1-q_2 \right| $. 
\end{lemma}	
\noindent \textbf{Proof of Lemma \ref{lemma:weightedmean}.} 
\noindent (a) In this case,  we can take 
\begin{itemize}
\item $b_1= \ldots = b_{T} \approx 0$; $b_{T+1}=\tilde{b}_{T+1}=c_2$; 
\item $a_1, \ldots, a_T$ can be arbitrary values that satisfy (\ref{cond2}); $a_{T+1}=d_1$, $\tilde{a}_{T+1}=d_2$.
\end{itemize}
This gives us $q_2-q_1=d_2-d_1$. Therefore, we should have $\max |q_2-q_1| \geq d_2-d_1$. At the same time each weighted average $q_1$ and $q_2$ has to belong to $[d_1,d_2]$, which is the range for $a_i$'s. Therefore, necessarily $\max|q_2-q_1| \leq d_2-d_1$. This implies that $\max|q_2-q_1|=d_2-d_1$. Note that if above we take $a_{T+1}=d_2$, $\tilde{a}_{T+1}=d_1$, then $q_2-q_1=d_1-d_2=-|d_2-d_1|$. 

\noindent (b) In this case, to evaluate the largest change in the weighted average we have to consider extreme situations. The first one is when $q_1=d_1$ and the $(T+1)$-th component in this average has the largest weight and changes to the other extreme $d_2$ in the new average $q_2$. 

This case can be described  as  
\begin{itemize}
\item $b_1= \ldots = b_{T}=c_1$; $b_{T+1}=\tilde{b}_{T+1}=c_2$; 
\item $a_1, \ldots, a_T=d_1$; $a_{T+1}=d_1$, $\tilde{a}_{T+1}=d_2$.
\end{itemize} 
This will give us $q_2-q_1=\frac{c_2 (d_2-d_1)}{T \cdot c_1+c_2}>0$.  

In the second extreme case $b_t$'s and $\tilde{b}_{T+1}$ are the same as above but $q_1=d_2$ and the $(T+1)$-th component in this average has the largest weight and changes to the the other extreme $d_1$ in the new average $q_2$. We obtain  $q_2-q_1=-\frac{c_2 (d_2-d_1)}{T \cdot c_1+c_2}<0$. These two extreme scenarios give us exactly the same  $|q_2-q_1| $. Thus,  $\max |q_2-q_1| =\frac{c_2 (d_2-d_1)}{T \cdot c_1+c_2} $. 

\noindent (c) Consider $b_i$, $i=1, \ldots, T+1$, and $\tilde{b}_{T+1}$ being  any values that satisfy (\ref{cond1}). Suppose $d_2=+\infty$.  Let $a_i$, $i=1, \ldots, T+1$, take any finite values while $\tilde{a}_{T+1}$ is very (arbitrarily) large.  This gives $q_1-q_2=-\infty$ and, thus, $|q_1-q_2|=+\infty$. Therefore, in this case $\max|q_1-q_2|=+\infty$. If, of course,  $\tilde{a}_{T+1}$ is taking a finite value while $a_T$ is very (arbitrarily) large, then  $q_1-q_2=+\infty$.

The case of $d_2$ finite but $d_1=-\infty$ is  analogous. $\blacksquare$

\vskip 0.05in 

\textbf{Extension of Lemma \ref{lemma:weightedmean}}: completely analogous results can be formulated for the case when the second average is $q_2=\sum_{i=1}^T \tilde{w}_i a_i$, where $\tilde{w}_i= {b_i}/{\sum_{j=1}^{T} b_j}$, $i=1, \ldots, T$.

Lemma \ref{lemma:weightedmean} and its extension are enough for us to evaluate the global sensitivity of an RDD estimator $\widehat{\theta}$.  The exact results on the global sensitivity depend on the type of kernel used in $\widehat{\theta}$. To better describe these facts, we differentiate among different types of kernels.  

First, we can have a  kernel $K(\cdot): \real \rightarrow \real^{+}$ \textit{with  a bounded support} which would mean that  there is a value ${u}_0>0$ such that $K(u)=0$ when $|u| > u_0$. 
For such kernels we define $\underline{K} \equiv \inf_{u \in (-u_0,u_0)} K(u)$. All other kernels are ultimately kernels with \textit{unbounded supports}. Uniform ($\underline{K}>0$), Epanechnikov, triangular ($\underline{K} =0$) are examples of kernels   with bounded supports whereas Gaussian and logistic  are examples of kernels with unbounded supports. 

For  bounded support kernels we can use use the notion a $K$-$h$-neighborhood defined next. For a given bandwidth $h$, we define  a $K$-$h$-neighborhood to the right (left) of $c$ as a set $[c,c+\Delta_{K,r}(h))$ ($(c-\Delta_{K,l}(h),c)$), where $\Delta_{K,r}(h)>0$ ($\Delta_{K,l}(h)>0$), such that $K(\frac{u-c}{h})>0$ if $u \in [c,c+\Delta_r(h))$ ($u \in (c-\Delta_{K,l}(h),c)$) and $K(\frac{u-c}{h})=0$ if $u \geq c+\Delta_r(h))$ ($u \leq c-\Delta_{K,l}(h),c)$) .  For kernels with unbounded supports, the $K$-$h$-neighborhood to the right (left) is $(c,+\infty)$ ($(-\infty,c)$). 

For kernels with a bounded support, let $\mathcal{Y}^{r}(h)$ ($\mathcal{Y}^{l}(h)$) denote the support of the distribution of $Y$ conditional on $X$ taking values in the  $K$-$h$-neighborhoods to the right (left). These supports are naturally approximated by 
\begin{equation} \label{notation:yplusminus} \mathcal{Y}^{r} = \lim_{h \downarrow 0} \mathcal{Y}^{r}(h) \text{ and } \mathcal{Y}^{l} = \lim_{h \downarrow 0} \mathcal{Y}^{l}(h)
\end{equation}	
that no longer depend on the bandwidth choice.\footnote{These limits are well defined as $\{\mathcal{Y}^{r}(h)\}$ and $\{\mathcal{Y}^{l}(h)\}$ are sequences of monotonically decreasing events when, without a loss of generality, $h$ decreases to zero in a monotonic fashion.} 
We will suppose that $\mathcal{Y}^{r}$ and $\mathcal{Y}^{l}$ are convex non-singleton sets. As further notations, we will use 
\begin{equation*}
\overline{Y}^{r} = \sup \mathcal{Y}^{r}, \quad   \underline{Y}^{r} = \inf \mathcal{Y}^{r},\quad 
\overline{Y}^{l} = \sup \mathcal{Y}^{l}, \quad \underline{Y}^{l} = \inf \mathcal{Y}^{l}.
\end{equation*}	

Our first result is when $K(\cdot)$ is  with a bounded support and continuous at the boundary. 
\begin{prop}
\label{prop:srd1} 
Consider  $\widehat{\theta}$, where $K(\cdot)$ is a kernel with a bounded support and $\underline{K}=0$. Suppose that for a data-driven choice of bandwidth $h=h(N)$, for any $N$ the  minimum number of observations in the $K$-$h$-neighborhood to the right (left) of $c$ is $m^{r}(N) \geq 1$ ($m^{l}(N) \geq 1$). 

\begin{enumerate}
\item[(a)] If $\mathcal{Y}^r$ and $\mathcal{Y}^l$ are bounded, then  the global sensitivity of $\widehat{\theta}$ is $\overline{Y}^{r}-\underline{Y}^{r} + \overline{Y}^{l}-\underline{Y}^{l}$ (and, thus, does not depend on $N$). 

\item[(b)] If at least one of $\mathcal{Y}^r$ and $\mathcal{Y}^l$  is unbounded, the global sensitivity of $\widehat{\theta}$ is $+\infty$. 
\end{enumerate} 
\end{prop}
\noindent \textbf{Proof of Proposition \ref{prop:srd1}.}
\noindent (a) The global sensitivity of  the  estimator is calculated by comparing the results of estimation for two datasets that differ only in on data point.  Thus, in RDD we need to consider the following situations:  (i) the new data point entesr a $K$-$h$-neighborhood of $c$ (and the old data point was outside of both $K$-$h$-neighborhoods of $c$); (ii) the new data point falls outside of both $K$-$h$-neighborhoods of $c$ (and the old data point was inside one of neighborhoods); (iii) the new data point remains in the same neighborhood; (iv) the new data point switches neighborhoods. 

Situations (i) and (ii) are analogous, and (iv) can be considered as a combination of two changes in (iii). Thus, it is enough to consider only (i) and (iv). We will use Lemmas \ref{lemma:weightedmean} and its extension with $c_2=\bar{K}$, where $\bar{K}$ denotes the supremum  value of $K(\cdot)$, and $c_1=\underline{K}$. 

(i) Suppose the new data point enters the $K$-$h$-neighborhood to the left of $c$ while the old data point was outside of both $K$-$h$-neighborhoods. By the extension of part (a) of Lemma \ref{lemma:weightedmean}, the maximum absolute change in the estimate is $\overline{Y}^{l}-\underline{Y}^{l}.$ Analogously, for the $K$-$h$-neighborhood to the right of $c$, the maximum absolute change in the estimate is
$\overline{Y}^{r}-\underline{Y}^{r}$. 

(iv) Suppose an observation moves from the left to the  right $K$-$h$-neighborhood of $c$. Since $\widehat{\theta}$ is the difference between the weighted means in the right  and left $K$-$h$ neighborhoods of $c$,  the described move affects both parts of the difference. 

From part (a) of Lemma \ref{lemma:weightedmean}, the maximum absolute change in the weighted average in the neighborhood to the right of $c$ is  $\overline{Y}^{r}-\underline{Y}^{r}$ and  this degree of change can be attained as a positive change (increase).  Similarly, the maximum absolute change in the weighted average for the left-hand side is $\overline{Y}^{l}-\underline{Y}^{l}$ and that this degree of change can be attained as a negative change (decrease). To obtain the maximum absolute changes for the difference in weighted means we have to look at the cases when these two weighted means change in opposite directions, which leads to the maximum change being $\overline{Y}^{r}-\underline{Y}^{r} + \overline{Y}^{l}-\underline{Y}^{l}$.  

The case when an observation moves from the right to the left $K$-$h$-neighborhood of $c$ is analogous. To sum up part (a), the global sensitivity is 
$ \overline{Y}^{r}-\underline{Y}^{r} + \overline{Y}^{l}-\underline{Y}^{l}.$ 

\noindent (b) Let, e.g., $\mathcal{Y}^r$ be  unbounded. Then part (c) of Lemma  \ref{lemma:weightedmean} will immediately give us   
$ +\infty$ global sensitivity when we  change one observation in the neighborhood to the right of $c$ by only changing its value of $y_i$ to an extreme value. $\blacksquare$ 

Our next case is of a kernel function with a bounded support and $\underline{K}>0$. For  simplicity, in the statement of  Proposition \ref{prop:srd2} we only indicate the rate of the global sensitivity. However,  the proof of the proposition gives an  exact expression for this sensitivity. 
\begin{prop}
\label{prop:srd2} 
Consider $\widehat{\theta}$ with $K(\cdot)$ being a kernel with a bounded support and $\underline{K}>0$. Suppose that for a data-driven choice of bandwidth $h=h(N)$, for any$N$ the minimum number of observations in the $K$-$h$-neighborhood to the right (left) of $c$ is $m^{r}(N) \geq 1$  ($m^{l}(N) \geq 1$).  

\begin{enumerate}
\item[(a)] If $\mathcal{Y}^r$ and $\mathcal{Y}^l$ are bounded, then  the global sensitivity of $\widehat{\theta}$ is proportional to $\frac{1}{\min\{m^{r}(N), m^{l}(N)\}}$. 

\item[(b)] If at least one of  $\mathcal{Y}^r$ or $\mathcal{Y}^l$ is unbounded, the global sensitivity of $\widehat{\theta}$ is $+\infty$. 
\end{enumerate} 
\end{prop}
 
\noindent \textbf{Proof of Proposition \ref{prop:srd2}.} Just like in Proposition \ref{prop:srd1}, the global sensitivity is 
determined by situation (iv) described in the proof of Proposition  \ref{prop:srd1}. For the exact expression for the global sensitivity we  rely on results  in Lemma  \ref{lemma:weightedmean} (and its extension)   but this time in part (b) of Lemma  \ref{lemma:weightedmean} we take $c_1=\underline{K}$ and $c_2=\overline{K}$. 

(a) Applying results of part (b) of Lemma \ref{lemma:weightedmean}, we obtain that  
when an observation from $K$-$h$ neighborhood to the left of $c$ moves to the neighborhood to the right of $c$, the largest absolute change in the estimator is 
$G_{LR} = \max\left\{S_{m^{l}(N)},S_{N-m^{r}(N)} \right\}$, where 
$$S_{m} \equiv \frac{\overline{K}\cdot (\overline{Y}^{l} - \underline{Y}^{l})}{m \cdot \underline{K} +\overline{K}} + \frac{\overline{K}\cdot (\overline{Y}^{r} - \underline{Y}^{r})}{(N-1-m) \cdot \underline{K} +\overline{K}}.$$
 When an observation moves from the right $K$-$h$ neighborhood of $c$ to the left, the maximum change in $\widehat{\theta}$ is   $G_{RL}=\max\{
 S_{N-1-m^{r}(N)},S_{m^{l}(N)-1}\}$. 

Thus, the global sensitivity can be concluded to be $\max\{G_{LR}, G_{RL}\}$ and is, clearly, of the rate $\frac{1}{\min\{m^{l}(N), m^{r}(N)\}}$. 

(b) If $\mathcal{Y}^r$ or $\mathcal{Y}^r$ is unbounded, then part (c) of Lemma  \ref{lemma:weightedmean} immediately gives us the infinite global sensitivity, just like in the proof of Proposition \ref{prop:srd1}. 
$\blacksquare$

Part (a) in Proposition \ref{prop:srd2} seemingly gives some hope of achieving a  situation when the global sensitivity may be going to zero as $N \rightarrow \infty$ if it can be ensured that $\min\{m^{r}(N), m^{l}(N)\} \rightarrow \infty$. This hope, however, is short-lived as it is possible to have realizations of samples $\{X_i\}_{i=1}^N$ such the number of observations in the  neighborhood to the right (left) is strictly less than $m^{r}(N)$  ($m^{l}(N)$). Indeed,
$\sum_{k=0}^{m^{side}(N)} \binom{N}{k} F_X(c)^{N-k} \left(1-F_X(c) \right)^k $ is the probability of fewer than $m^{side}(N)$  observations to the right (left) of  $c$ when $side=r$ ($side=l$). 
This probability is strictly positive when $c$ is an interior point of the support of $X$. Thus, the global sensitivity  has to be taken as bounded away from 0 as $N \rightarrow \infty$ in the case of the kernel with a bounded support and $\underline{K}>0$. 

In our final case -- that of a kernel with  an unbounded support, -- the results are completely analogous to those in Proposition \ref{prop:srd1}.  

\subsection{Global sensitivity of the local linear estimator}


First, consider the sharp design.
Since the local linear estimator effectively considers observations whose running variable values  are in a small neighborhood around $c$, we employ (\ref{notation:yplusminus})  as approximations of the support for the treatment outcome. 

As we know, 
$$
\widehat{\alpha}_R  = \overline{y}_R-\frac{\overline{x}_R-c}{\sum_{i=1}^N (x_i q_i -\overline{x}_R)^2\cdot 1(c \leq x_i)} 
\cdot \sum_{i=1}^N (q_i x_i-\overline{x}_R) q_i y_i\cdot 1(c \leq x_i),
$$
where $q_i=K\left(\frac{x_i-c}{h_N}\right)$, $\overline{y}_R = \frac{\sum_{i=1}^N q_i y_i 1(c \leq x_i)}{\sum_{i=1}^N q_i 1(c \leq x_i)}$, $\overline{x}_R = \frac{\sum_{i=1}^N q_i x_i 1(c \leq x_i)}{\sum_{i=1}^N q_i 1(c \leq x_i)}$. An analogous formula applies to $\widehat{\alpha}_L$. 

We can show that the global sensitivity is infinite even if $\mathcal{Y}^r$ or $\mathcal{Y}^r$ are bounded. Let us show that the maximum change in $\widehat{\theta}_{S,LocLin}$ is infinite when one observation in the $K$-$h$ neighborhood to, e.g.,  the right of $c$ changes but stays within that neighborhood. Consider a dataset where the first $T\leq N$ (and only those) observations are in that neighborhood. Consider a realization of a new dataset when only $T$-th observation changes its value.  Suppose we have the following realized data: 
\begin{align}
x_i&= c+\Delta_N, \quad i=1, \ldots, m^{r}(N)-1 \label{proofLL_ex1}\\
x_{T} &= c+u_0h_N-\Delta_N, \quad 	x_{T} '= c+\Delta_N-\Delta_N\gamma, \label{proofLL_ex2}
\end{align}	
for some $0<\Delta_N<u_0 h_N$ and $0<\gamma<1$. For a kernel $K(\cdot)$ with a bounded support $u_0>0$ is the value such that $(-u_0,u_0)$ is the support of this kernel. If $K(\cdot)$ has an unbounded support, then we can take $u_0$ to be a very large positive number. In either case, we can take 
\begin{equation*}
q_i \approx K(0), \quad i=1, \ldots, m^{r}(N)-1, \quad 
q_{T} \approx \underline{K}, \quad 	q_{T} '= K(0).
\end{equation*}	

Suppose that $y_{T}=y_{T}'$. Then 
\begin{equation*}
\widehat{\alpha}_R  \approx \overline{y}_R+\Delta_N\left(1-\frac{\gamma }{T}\right)\times 
\frac{{\Delta_N \gamma}\sum_{i=1}^{T-1} y_i/T +  y_T ({T-1}) \Delta_N \gamma /T}
{{(T-1)\Delta_N^2 \gamma^2}/{T^2}+ ((T-1) \Delta_N \gamma)^2/T^2}.  
\end{equation*}	
For fixed $T$, $h_N$, $\Delta_N$, it is possible to have $\gamma \downarrow 0$, in which case we have that 
$|\widehat{\alpha}_R' - \widehat{\alpha}_R|\rightarrow \infty.$
Since there are no changes in $\widehat{\alpha}_L$, we conclude that  the global sensitivity is infinite.  

Note that when the kernel either has an unbounded support or has a bounded support with $\underline{K}=0$, then even without using $\gamma \downarrow 0$, we can establish that the global sensitivity is bounded away from zero for any $N$, using techniques similar to those in Proposition \ref{prop:srd1}.

When the support of $Y|X$ in the neighborhood  of $c$  is unbounded, then the global sensitivity is obviously infinite, which can be shown by just changing one value of $Y_i$ only.

\vskip 0.05in 

Similar conclusion would be drawn for the the fuzzy design where local linear estimation would have to use the IV version of the estimator above. 

\subsection{Implication for  regular smooth DP RDD estimators} 

The results on the global sensitivity for nonparametric regression at the boundary and local linear estimators imply that additive mechanisms which build directly on these estimators (thus, which have the form $\widehat{\theta}+a(\nu_N)$) will have the variance of $a(\nu_N)$ that does not diminish to zero if privacy constraints remain bounded from above (variance of $a(\nu_N)$ decreasing to 0 is incompatible with the DP guarantees). Under informational asymmetry, a researcher may have to consider a whole set of such mechanism noise random variables $a(\nu_N)$ available leading to the weak limit which is random set and whose distribution is driven by the distribution limits of $a(\nu_N)$. The same issues arise for smooth regular DP estimator in the form $\widehat{\theta}+a(\nu_N)+\Delta_N$. 

We can consider other smooth DP mechanisms in this setting -- e.g. those based on the application of the post-processing theorem mentioned in the introduction. 
Whenever we have a WLS-type estimator, to ensure e.g., $(\varepsilon_N, \delta_N)$-differential privacy, noise components 
would be added individually to the numerators and the denominators of the ratios in our WLS estimator. 
Then the post-processing theorem of DP (\cite{dwork2014algorithmic}) would imply $(\varepsilon_N, \delta_N)$-differential privacy of the original WLS estimator. Qualitatively, our conclusions with respect to nonparametric regression at the boundary and  local linear estimator would remain the same.

Thus, our general conclusion for a class of smooth differentially private nonparametric RDD estimators is the failure to identify the treatment effect in the limit of statistical experiments. If we add other covariates to our estimation,  the qualitative conclusions on the lack of identifiability of the parameter of interest will remain the same.   

\subsubsection{Impact of DP on validity checks and graphical analyses in RDD}
\label{sec:rdd_additional_short}

In regression discontinuity designs (RDD), validity depends on confirming that any observed discontinuities at the treatment cutoff truly reflect causal effects rather than other structural breaks or data manipulation. Two main diagnostic checks are typically used: (i) \emph{placebo tests} with pre-treatment covariates or pre-treatment outcomes, and (ii) \emph{tests for manipulation} of the forcing variable. Differential privacy (DP) introduces significant challenges for both.

\paragraph*{Placebo Tests} 
Placebo tests examine whether covariates unrelated to the treatment, or outcomes measured before treatment, also exhibit a discontinuity at the cutoff. These tests usually rely on $t$-statistics. Under DP, the tests themselves must satisfy $(\tilde{\varepsilon}_N, \tilde{\delta}_N)$-privacy. However, because the sensitivity of an RDD estimator does not diminish with sample size, the additional DP noise required to protect privacy quickly dominates the statistic’s natural variation. Even if critical values are adjusted for this noise, the test loses power eventually reflecting mostly the randomness introduced by the privacy mechanism rather than any true effect. As a result, differentially private placebo tests often yield unreliable or trivial inference.

\paragraph*{Manipulation Tests} 
The density continuity test proposed by \cite{McCrary2008}, designed to detect manipulation of the forcing variable, faces similar limitations. Local density estimators are highly sensitive, and in any smooth regular DP version of the test statistic, the privacy mechanism noise overwhelms the meaningful signal. Consequently, DP-compliant versions of the manipulation test have very low power and provide limited evidence about potential sorting or manipulation near the threshold.

\paragraph*{Graphical Analyses} 
Visual checks such as binned scatterplots of outcomes, covariates, or the forcing variable are important in an RDD analysis. However, constructing DP histograms or bin means requires either randomizing bin placement or injecting substantial noise. This disrupts the alignment of bins with the cutoff and can obscure or distort discontinuities, undermining the purpose of the visualization. Even with large samples, the added DP noise remains non-negligible, often creating artificial patterns or masking real ones.

\end{document}